\documentclass[%
      aps,
      prb, %jcp,
      twocolumn,
      superscriptaddress,
      showpacs]{revtex4-1}
      
\usepackage[utf8]{inputenc}
\usepackage[T1]{fontenc}
\usepackage{dcolumn}
\usepackage{bm}					% Bold Greek letters in math mode.
\usepackage{graphicx}
\usepackage{amsmath, amsfonts, amssymb, mathtools}	
\usepackage{latexsym} 				% Additional symbols. 
\usepackage{xcolor}
\usepackage[caption = false]{subfig}
\usepackage[breaklinks, linkcolor = black]{hyperref}
\usepackage{tikz}
\usepackage{tabularx}
\usepackage{braket}
\usepackage{diagbox}
\usepackage{ulem} 
\usepackage{textcomp}
\usepackage{soul}
\usepackage[mathlines]{lineno} % Enable numbering of text and display math
\usepackage{mathptmx}
\usepackage{blindtext}
\usepackage{booktabs,eqparbox}
\usepackage{siunitx}
\usepackage{physics}

\newcommand{\black}[1]{\textcolor{black}{#1}}
\makeatletter
\let\@fnsymbol\@fnsymbol@latex
\@booleanfalse\altaffilletter@sw
\makeatother

\begin{document}

\title{\black{Mixed excitonic nature in water-oxidized BiVO$_4$ surfaces with defects}}

%%%%%%%%%%%%%%%%%%%%%%%%%%%%%%%%%%%%%%%%%%%%%%%%%%%
% 
\author{Rachel Steinitz-Eliyahu}

\affiliation{Department of  Molecular Chemistry and Materials Science, Weizmann Institute of Science, Rehovot 7610001, Israel}

\author{Daniel \surname{Hernang\'{o}mez-P\'{e}rez}}

\affiliation{Department of  Molecular Chemistry and Materials Science, Weizmann Institute of Science, Rehovot 7610001, Israel}

\author{Franziska S. Hegner}
\affiliation{Institut Catal\`a d'Investigaci\'o Qu\'imica (ICIQ), The Barcelona Institute of Science and Technology (BIST), Avda. Pa\"isos
Catalans, 16, 43007 Tarragona, Spain.}

\author{Pavle Nikačević}
\affiliation{Institut Catal\`a d'Investigaci\'o Qu\'imica (ICIQ), The Barcelona Institute of Science and Technology (BIST), Avda. Pa\"isos
Catalans, 16, 43007 Tarragona, Spain.}

\author{N\'uria L\'opez}
\affiliation{Institut Catal\`a d'Investigaci\'o Qu\'imica (ICIQ), The Barcelona Institute of Science and Technology (BIST), Avda. Pa\"isos Catalans, 16, 43007 Tarragona, Spain.}

\author{Sivan Refaely-Abramson}

\affiliation{Department of  Molecular Chemistry and Materials Science, Weizmann Institute of Science, Rehovot 7610001, Israel}

\email[Corresponding author:]{sivan.refaely-abramson@weizmann.ac.il}

%%%%%%%%%%%%%%%%%%%%%%%%%%%%%%%%%%%%%%%%%%%%%%%%%%%%%%%%

\keywords{BiVO\textsubscript{4}, water splitting, electron transfer, GW}

\begin{abstract}
 BiVO$_4$ is a promising photocatalyst for efficient water oxidation, with surface reactivity determined by the structure of active catalytic sites. Surface oxidation in the presence of oxygen vacancies induces electron localization, suggesting an atomistic route to improve the charge transfer efficiency within the catalytic cycle. In this work, we study the effect of oxygen vacancies on the electronic and optical properties at BiVO$_4$ surfaces upon water oxidation. We use density functional theory and many-body perturbation theory to explore the change in the electronic and quasiparticle energy levels and to evaluate the electron-hole coupling as a function of the underlying structure. We show that while the presence of defects alters the atomic structure and largely modifies the wavefunction nature, leading to defect-localized states at the quasipatricle gap region, the optical excitations remain largely unchanged due to substantial hybridization of defect and non-defect electron-hole transitions. Our findings suggest that defect-induced surface oxidation supports improved electron transport, both through bound and tunable electronic states and via a mixed nature of the optical transitions, expected to reduce electron-hole defect trapping.   
\end{abstract} 
 
\maketitle

Metal oxide semiconductors serve as promising photoanodes for solar-driven water splitting processes~\cite{jiang2017photoelectrochemical}. Bismuth vanadate (BiVO\(_4\)) is a famous example of exceeding interest, being relatively stable, non toxic, and long-lasting, with absorption well within the solar spectrum \cite{jia2012facile, kudo1998photocatalytic,sharp2017bismuth,tan2017alternative,tolod2017recent,zhao2011electronic,wiktor2017comprehensive,das2017investigation,wang2021exciton}, and with suitable electronic properties for efficient oxygen evolution reaction (OER)~\cite{sharp2017bismuth,tolod2017recent, cooper2014electronic, park2013progress}. Despite these appealing properties, electron conductivity and carrier transport through BiVO\(_4\) surfaces is found to be relatively low~\cite{abdi2013origin,rettie2013combined,seo2018role}, greatly limiting its use in practical applications. This low conductance is typically attributed to fast charge recombination and polaronic interactions~\cite{kim2015simultaneous, park2011factors}, strongly coupled to surface structure~\cite{wang2018new,hammes2021integration,hu2018anisotropic,li2019crystallographic, wu2021unveiling,lardhi2020significant,kahraman2020fast,pavle} and the conditions in which it was prepared~\cite{hammes2021integration,li2017first, wang2020role, das2017investigation}. 
A broadly-explored pathway to improve BiVO\(_4\) functionality is through electronic doping upon element substitution and the introduction of surface vacancies near the interacting BiVO\(_4\) surface~\cite{li2015surfactant,kim2015simultaneous,wang2018new,fernandez2020role}. These induce modified electronic densities and allow controllable electron mobility and electron-hole recombination rates~\cite{rettie2013combined, luo2011solar, abdi2013efficient, liang2011highly}. 

Oxygen vacancies are typical intrinsic defects in BiVO\(_4\) that can also be generated and controlled via external treatments, such as hydrogen annealing and nitrogen flow~\cite{cooper2016role, kim2015simultaneous,daelman2020quasi}. By acting as $n$-type donors, these defect hold the promise to significantly increase the water oxidation reaction yield ~\cite{seo2018role,hegner2019versatile,ulpe2018f,wang2020role,hu2019bivo}. From an electronic structure point of view, oxygen vacancies introduce localized electronic states, allowing tunable electronic bandgaps and suggesting enhancement of the separation between photogenerated electrons and holes and improved absorption cross-section of the visible light ~\cite{kim2015simultaneous, seo2018role, yin2011doping, hegner2019versatile}. On the other hand, localized defect states are also considered as active electron-hole recombination centers, which can trap the photogenerated energy carriers- namely, excitons- and consequently reduce the electronic conductivity and the electron and energy transfer efficiency associated to it~\cite{zhao2018clarifying, yang2013new, wang2020role,liu2020hole,selim2019impact}. The role of oxygen vacancies in photogenerated carrier transport thus remains controversial, and a comprehensive understanding of the involved defect-induced phototransport mechanisms is still lacking.

Recent density functional theory (DFT) studies found that a structurally-stable split oxygen vacancy, in which the vacancy is shared between two neighboring vanadium atoms in the form of a V--O--V bridge, is pivotal to the catalytic reaction that produces molecular oxygen upon water adsorption at the thermodynamically-stable (001) BiVO\(_4\) surfaces~\cite{hegner2019versatile,pavle}. These studies showed that the change in surface structure and the underlying chemical bonding due to oxygen vacancies strongly depend on the adsorbate. In particular, main steps within the OER involve either a peroxo bridge between the Bi and V atoms at the surface, or a surface oxo group evolving into an OOH group. The electronic structure associated with the defects is thus expected to vary due to significant surface-structure modifications within the various steps in the catalytic reaction. The electronic states and the electron-hole binding associated with the vacancies within the catalytic reaction are hence non-trivial, and a predictive assessment of the associated electronic and excitonic fine structure as a function of these structural modifications is required.

\begin{figure}[h]
  \centering
  \includegraphics[width=0.465\textwidth]%
  {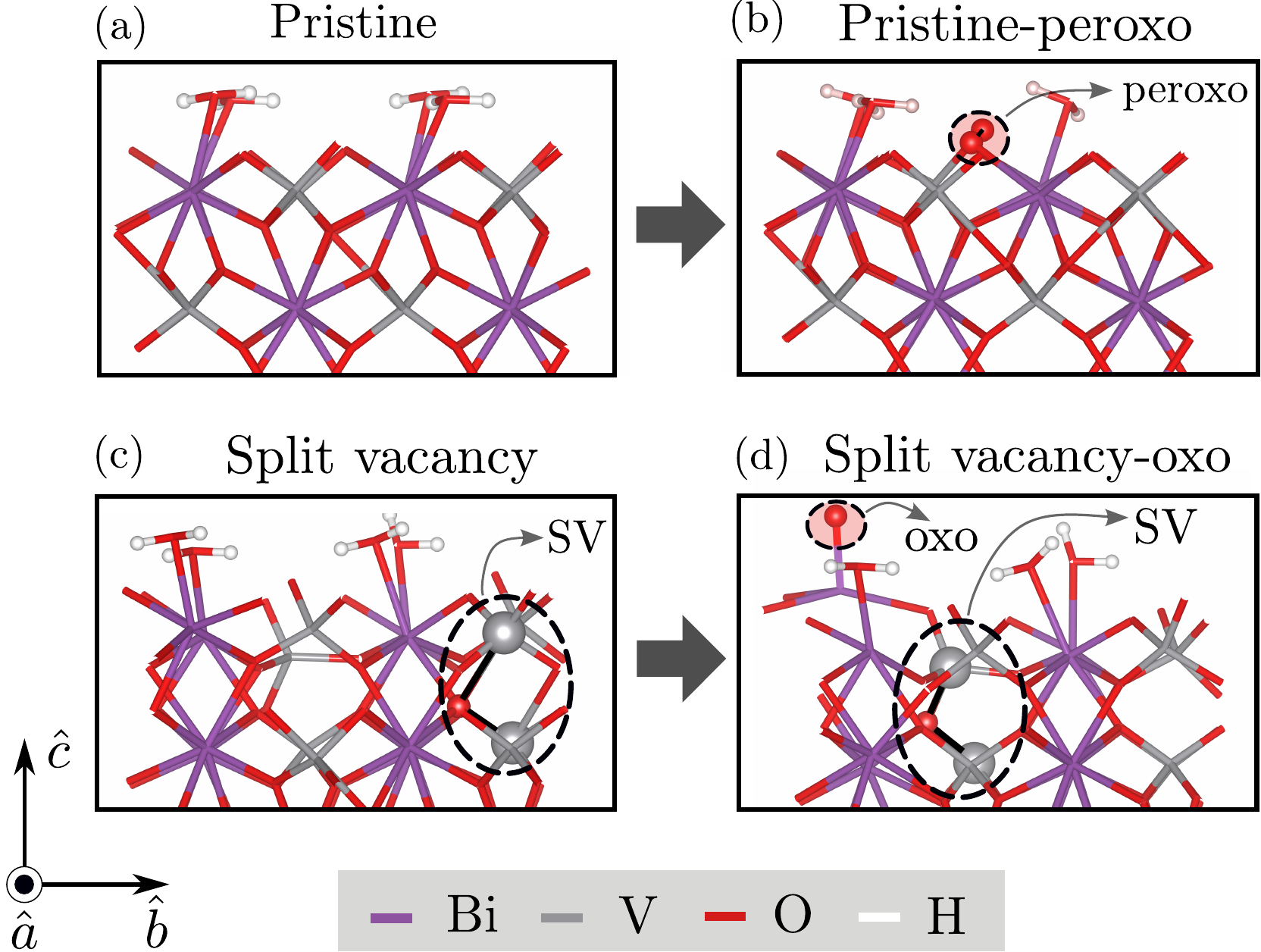}% picture filename
  \caption{Four  BiVO\textsubscript{4} surface structures upon water adsorption, studied in this work: (a) a pristine surface, with full water coverage; (b) a pristine-peroxo surface, in which one water molecule per repeating cell went through oxidation; (c) a surface including a split vacancy with full water coverage; and (d) a split vacancy-oxo surface, in which one water molecule in a surface including a split vacancy went through oxidation. Oxidized water molecules are marked by red circles; black arrows represent the structural change upon oxidation. The split vacancy, appears as a V--O--V bridge, is marked with a black dashed circle and denoted by `SV'. Each structure represents the repeating unit cells in the $\hat{a},\hat{b}$ direction; the $\hat{c}$ direction contains six layers as well as additional vacuum, and is shown in the SI for the case of the pristine system.  
  }
    \label{structures}
\end{figure}

In this work, we study the effect of oxygen vacancies on the electronic and excitonic properties in BiVO\(_4\) upon surface water oxidation. We use DFT and many-body perturbation theory within the GW and GW-Bethe-Salpeter equation (GW-BSE) approximation to calculate the quasiparticle energies and the electron-hole coupling for representative structures found to be stable during the water oxidation stage in the catalytic cycle. We explore how the introduction of oxygen vacancies and the charge localization associated to it influence these properties. Our results show defect-localized states nearby the valence region, associated with the modified underlying chemical bonding, with quasiparticle bandgap largely unchanged upon the inclusion of surface defects. This leads to largely mixed exciton states, hybridizing transitions between defect and non-defect bands. We find that due to this mixing, and surprisingly, the presence of defects does not alter the exciton binding energy in the examined systems. 
%In other words, the low lying excitons are not primarily formed from defect states. 
%Our findings suggest that while water oxidation in the presence of oxygen vacancies induces localized electronic states associated with the modified underlying chemical bonding, the associated electron-hole coupling is complex, and the induced changes in the exciton binding are small. 

The examined systems are shown in Fig.~\ref{structures}. Our focus is on the (001) facet of monoclinic scheelite, which is considered to be thermodynamically stable and thus the active polymorph of the BiVO$_4$ surface. The calculated unit cell includes six layers, each containing four Bi and four V atoms, with a vacuum of $14$~\AA\, separating repeating cells along the $\hat{c}$ direction (see SI). We considered four structures occurring within two oxidation pathways, recently suggested to play a key role within the photocatalytic cycle~\cite{pavle}:  a pristine BiVO$_4$ surface  (Fig.~\ref{structures}a); the same system after one water molecule per cell went through oxidation, resulting in a surface-peroxo group (Fig.~\ref{structures}b); a BiVO$_4$ surface with a subsurface split vacancy (SV) (Fig.~\ref{structures}c); and the same system after one water molecule per cell went through oxidation in the presence of a vacancy, resulting in an surface-oxo group (Fig.~\ref{structures}d). 

In the following, we present the calculated quasiparticle and excitonic properties associated with these structures, and examine the electronic distribution and electron-hole coupling as a function of oxidation with and without the subsurface defects. To compute the single-particle electronic structure and the wavefunctions of the examined surfaces, we employ density functional theory (DFT) within the Perdew-Burke-Ernzerhof (PBE) approximation~\cite{Perdew1996} for the exchange-correlation functional including spin-orbit coupling, using the \textsc{Quantum Espresso} package \cite{Giannozzi2009}. The resulting DFT energies and wavefunctions are then used as a starting point for our many-body perturbation theory calculations, performed within the BerkeleyGW package \cite{deslippe2012berkeleygw}. We compute quasiparticle energy corrections applying the G$_0$W$_0$ approach within the generalized plasmon-pole model for the frequency dependence of the dielectric screening \cite{Hybertsen1986}. Following standard converging procedures, we consider a $30$~Ry screening cut-off and a total of $3500$ electronic bands, among them $1280$ being occupied. Due to the large size of the explored repeating cells, we sample the reciprocal space with a relatively coarse uniform \textbf{k}-point grid of $2\times2 \times 1$. To account for electron-hole coupling and excitonic properties, we employ the Bethe-Salpeter equation (BSE) formalism within the Tamm-Dancoff approximation ~\cite{Rohlfing1998, *rohlfing2000electron}, while considering $14$ valence (occupied) bands and $16$ conduction (empty) bands in the coupling matrices (see full computational details in the SI).

\begin{figure}
    % \centering
    % \hfill
    \centering
     \includegraphics[width=0.49\textwidth]{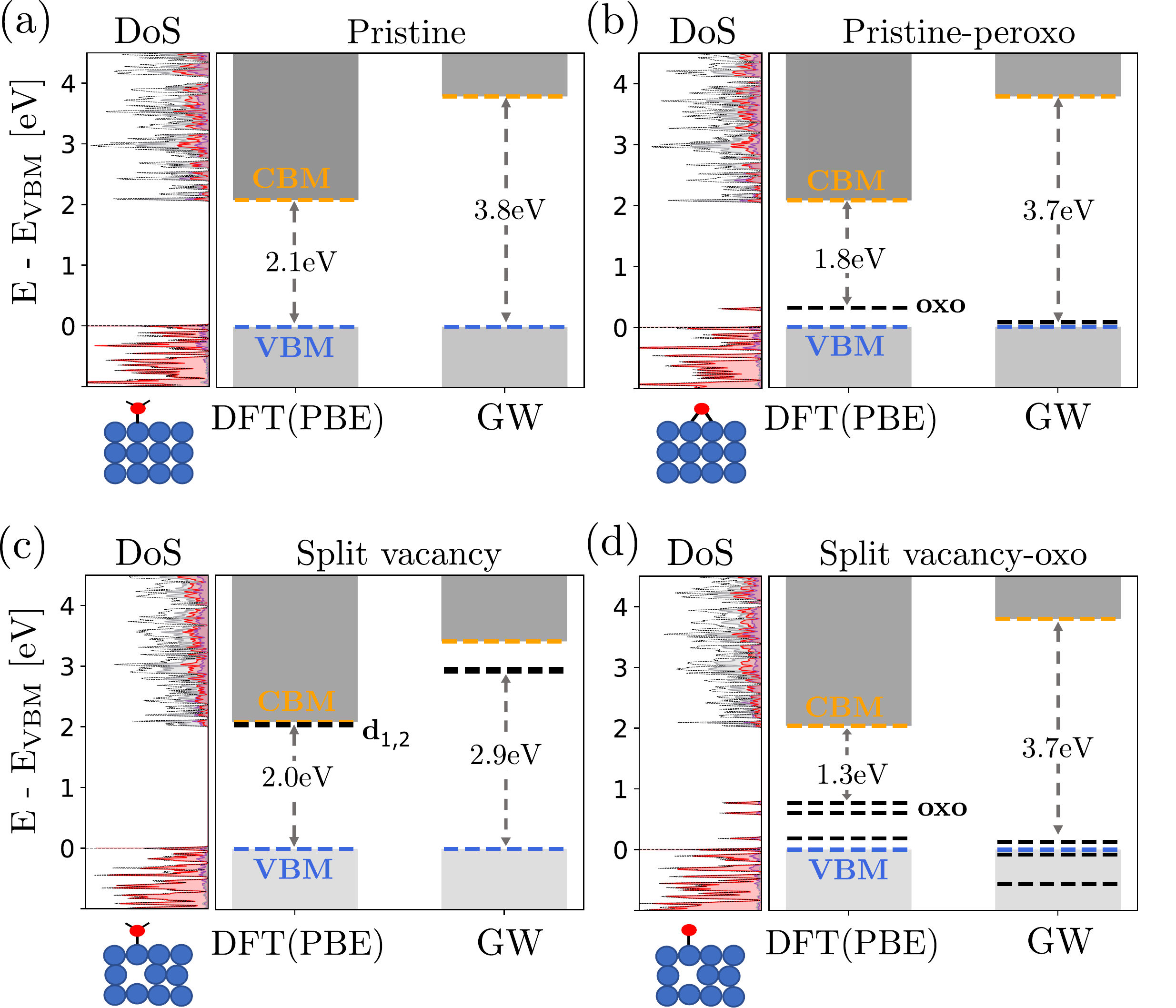}%
     \caption{Calculated DFT and GW electron and quasiparticle energies and bandgaps, for the (a) pristine, (b) pristine-peroxo, (c) split vacancy and (d) split vacancy-oxo systems at the $\Gamma$ point. 
     Pristine-like CBM and VBM states are indicated in orange and blue dashed lines, respectively, and additional oxo and defect energy levels are shown as black dashed lines.
     The projected density of states of each of the surface structures onto the atomic contributions is also shown (violet: Bi, grey: V, red: O, white: H), with the dashed line corresponds to the total density of states.}
         \label{eqp}
 \end{figure} 
 
Fig.~\ref{eqp} shows the computed DFT electronic energies and the GW quasiparticle energies, as well as the associated bandgaps, for the pristine, pristine-peroxo, split vacancy, and split vacancy-oxo surface structures, respectively. Here we present the calculated values at the $\Gamma$ point, since the band dispersion is small in the examined systems (full DFT bandstructures are given in the SI). Blue and orange colors represent the VBM (valence band maximum) and CBM (conduction band minimum) onsets, respectively. Black lines represent defect- and surface-oxidized states at the gap region. For the pristine system (Fig.~\ref{eqp}a), the GW opens up the DFT (PBE) gap significantly (by 1.7~eV), to a value of 3.8~eV. This GW gap is slightly larger than the ones reported before for the bulk system, of 3.4--3.6~eV~\cite{wiktor2017comprehensive}. The gap increase can be expected due to the reduced dielectric screening beyond the surface~\cite{qiu2016screening}. Upon water oxidation and the formation of a surface peroxo group (Fig.~\ref{eqp}b), the DFT results show an additional occupied in-gap state, which is shifted down to the VBM onset when including GW  quasiparticle corrections. 

In the presence of split vacancies (Fig.~\ref{eqp}c), the defect states marked as $d_{1,2}$ appear in the gap region. In this structure the GW gap of 2.9~eV is significantly smaller than in the pristine case. We associated this gap reduction to a partial state charging at the CBM region, responsible for the instability of this structure~\cite{pavle}. Notably, upon oxidation of the defect surface (Fig.~\ref{eqp}d), three occupied in-gap states, largely localized on the oxo group, appear at the valence edge region. The quasiparticle GW gap obtained is similar to the pristine and pristine-peroxo cases. These results suggest that the presence of oxygen vacancies induces localized states in addition to the pristine-like ones, which change their nature upon surface water oxidation. As we show in the following, these changes in the quasiparticle spectra result from the involved modifications in the chemical bonding around the missing atoms.

\begin{figure}
  \centering
  \includegraphics[width=0.5\textwidth]%
  {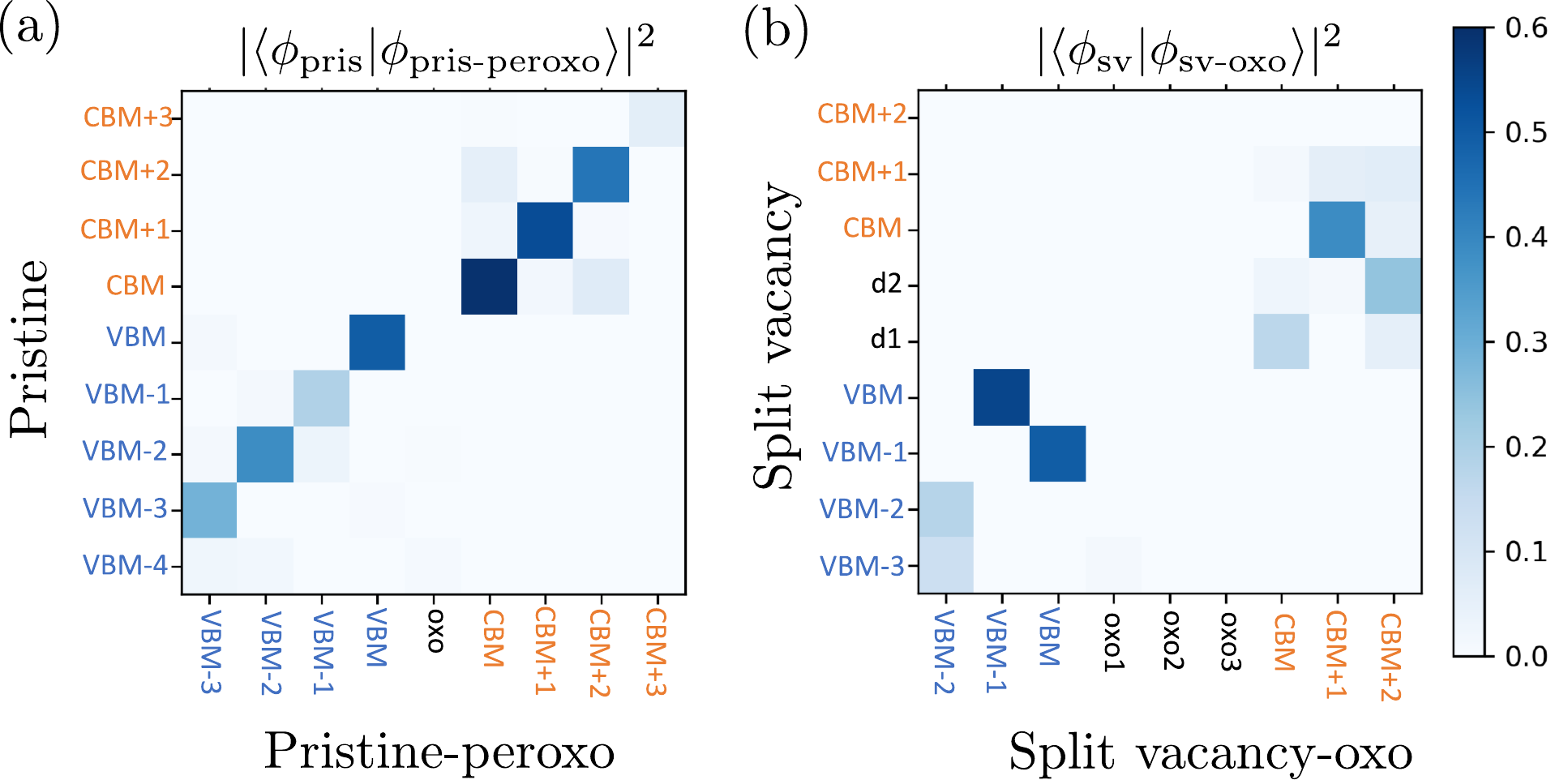}% picture filename
   % \includegraphics[width=0.5\textwidth]%
  %{figures/twopanels.pdf}% picture filename
  \caption{{(a)} Calculated coupling matrix elements $S_{AB}=\abs{\bra{\phi_{A}}\ket{\phi_{B}}}^2$ between DFT electronic states of the pristine and pristine-peroxo systems, where 0 shows no overlap and 1 shows complete overlap. {(b)} Same as in (a) for the comparison between split vacancy and split vacancy-oxo systems.}  
  %% DH: S-matrix is ambiguous, I would call this overlap matrix as S-matrix has a different meaning
  
    \label{S_matrix}
\end{figure}

To further quantify the defect- and oxidation-induced change in the electronic structure, we examine the wavefunction coupling before and after water oxidation, for the two pristine and the two split-vacancy structures. For this, we evaluate the coupling matrix elements between the computed DFT wavefunctions of each two structures. We define the overlap matrix $S_{AB}=\abs{\bra{\phi_{A}}\ket{\phi_{B}}}^2$, where $A$ and $B$ are different systems, and $\phi$ is an electronic Kohn--Sham state in a planewave basis set. As anticipated above, it is enough to only discuss the $S$ matrix at the $\Gamma$-point, as the other k-points show a similar behavior due to the small band dispersion in the Brillouin Zone (see SI). Fig.~\ref{S_matrix}a shows the computed overlap between the pristine and pristine-peroxo structures, $S=\abs{\bra{\phi_{\textnormal{pris}}}\ket{\phi_{\textnormal{pris-peroxo}}}}^2$. For this case, we find substantial coupling between the two systems - before and after surface oxidation - at the valence and conduction areas. As expected, the additional peroxo-localized in-gap states have negligible overlap with any of the pristine bands, suggesting that the local modification due to the surface oxidation and the peroxo formation is small.

In contrast, in the presence of sub-surface split-vacancy defects, this picture becomes more complicated. The computed overlap between the split vacancy and split vacancy-oxo structures, $S=\abs{\bra{\phi_{\textnormal{sv}}}\ket{\phi_{\textnormal{sv-oxo}}}}^2$ (Fig.~\ref{S_matrix}b), shows that while few bands around the valence and conduction area remain of similar nature, the defect states largely change their nature before and after surface oxidation. As a result, we find negligible overlap between defect states with water surface adsorbates (fig.~\ref{eqp}c) and defect states with oxidized adsorbates (fig.~\ref{eqp}d), namely there are significant changes in the wavefunction nature due to the surface oxo-bonding and the structural changes associated to it.
We further note that the overlap matrix elements vanish almost completely when comparing between the pristine and the split vacancy structures, and between the pristine-peroxo and the split vacancy-oxo structures (see SI). We hence find that the presence of oxygen vacancies leads to changes in the underlying bonding, which completely varies the electronic wavefunctions; and surface oxidation changes the surface bonding in the presence of defects, leading to the observed modifications in the electronic and quasiparticle spectra.

\begin{figure*}
     \centering
     \hfill
     
    \centering
    \includegraphics[width=0.8\linewidth]{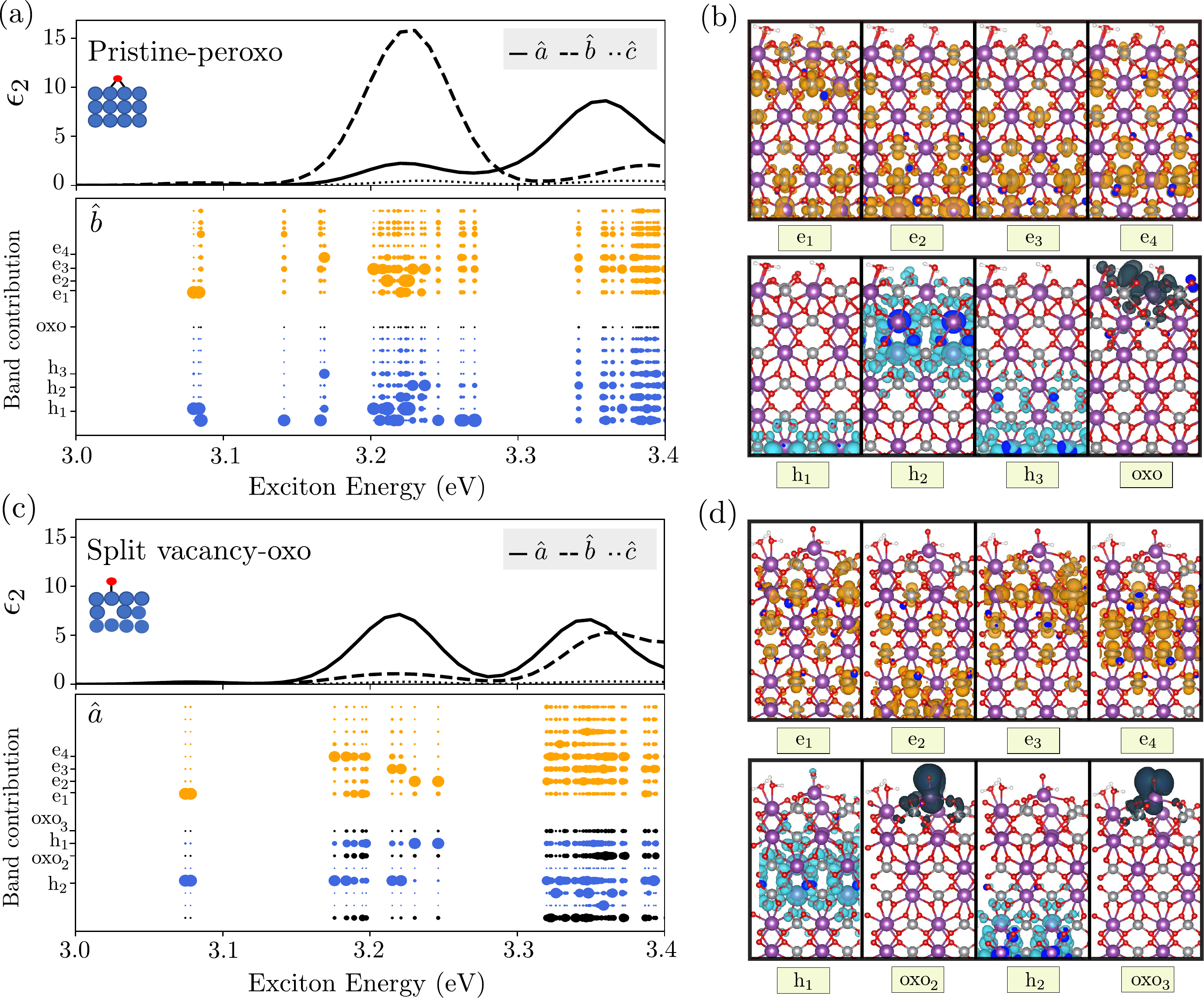}
  %%% DH: I have added a little bit more space between the plot and the x-axis label 
     \caption{
     (a) Calculated GW-BSE absorption spectra for the three main polarization directions, as well as electron-hole transitions contributing to the absorption peaks, for the pristine-peroxo system. Contributions from occupied (hole) bands are shown in blue, from unoccupied (electron) bands in orange, and from localized peroxo and oxo bands in black. Each dot in the lower panel represents the band contribution to the exciton as a function of the  excitation energy, weighted by the oscillator strength at the dominating polarization direction  ($\hat{b}$). 
     (b) Representative wavefunction distributions corresponding to the electron and hole Kohn-Sham states with the largest contribution to the low-energy absorption peaks.
     (c) - (d) Same as in (a) - (b) for the split vacancy-oxo system. The exciton contributions are weighted with oscillator strength at the $\hat{a}$ polarization direction, which is the dominating one in this structure. 
     }
         \label{excitons}
 \end{figure*}

Having identified the change in electronic states and quasiparticle energy levels upon surface water oxidation and in the presence of defects, we now turn to compute the electron-hole coupling and the associated optical and excitonic properties.
Fig.~\ref{excitons} shows the obtained GW-BSE absorption spectra of the two oxidized surfaces with and without sub-surface defects, pristine-peroxo and split vacancy-oxo, upon optical excitation with light polarized at the three main crystal directions. The absorption peaks are further analysed through the electron-hole transitions composing them.  
We first note that in both structures, the absorption oscillator strengths at the $\hat{a}$ and $\hat{b}$ polarization directions are much larger than at the $\hat{c}$ direction, suggesting that in-plane excitations dominate the spectra, associated with the layered coupling in the explored structures. However, while the lowest excitation peak in the pristine-peroxo case is at the $\hat{b}$ polarization direction, in the split vacancy-oxo case it is at the $\hat{a}$ direction, serving as another signature of the significant underlying structural changes involved upon the presence of defects.

For the pristine-peroxo structure, Fig.~\ref{excitons}a, the lowest bright excitation is at 3.08~eV, and the low-energy absorption peak is at 3.23~eV. These excitation energies are in good agreement with experimental findings, where the optical gap is found at $\sim$3~eV~\cite{stoughton2013adsorption,cooper2015indirect, wu2021unveiling}. 
The differences between the computed quasiparticle and optical gaps point to a relatively large exciton binding energy, at the order of $0.6$ eV, suggesting strong electron-hole binding. We note that this energy difference does not include lattice relaxation processes, which can largely reduce the bandgap in these materials~\cite{payne2011nature,stoughton2013adsorption,cooper2014electronic,cooper2015indirect}. 
The computed excitation energies for the non-oxidized pristine case are very similar (see computed absorption in the SI), and are in good agreement with previous calculations for the bulk pristine system~\cite{wiktor2017comprehensive}. Such agreement between surface and bulk structures can be explained by the local nature of the electron-hole interactions, leading to small long-range surface effects on the computed excitation energies. We note however that this points to larger exciton binding energy at the surface compared to the bulk, as expected, due to an increase in the quasiparticle gap while the optical gap remains similar. 

 The exciton coefficients for each excitonic state $S$, \(A^S_{vc\bf{k}}\), quantify the coupling between an electron at band $c$ and a hole at band $v$, at the k-point $\bf{k}$. Electron-hole transitions composing the absorption spectra of the pristine-peroxo system are shown below the corresponding absorption energies in Fig.~\ref{excitons}a. Dot size represents the relative magnitude of the normalized exciton contribution, $\sum_{c\bf{k}}|A^S_{vc\bf{k}}|^2$ for unoccupied ($c$) bands and $\sum_{v\bf{k}}|A^S_{vc\bf{k}}|^2$ for occupied ($v$) bands, summed over all k-points and scaled with the oscillator strength at the dominating polarization direction, so that only bright peaks are shown (contributions weighted by the other polarization directions are shown in the SI). This analysis allows us to quantify the contribution from each electron/hole state (orange/blue dots, respectively) to each one of the computed excitons. Black dots represent localized and occupied surface oxidation states, associated to the dashed black states of Fig.~\ref{eqp}b. The electron/hole wavefunctions with the most significant contributions to low-lying excitons are shown in Fig.~\ref{excitons}b. Notably, the lowest bright exciton around 3.1~eV is mainly composed of coupling between the conduction electron band, $e_1$, and a hole band below the valence region, $h_1$. We note that this hole state is mainly localized at the bottom layer and thus may have increased delocalization when more layers are included, bringing it closer to the higher-energy hole states at the valence region. The peak around 3.2~eV already includes multiple electron-hole transitions, with additional contributions from occupied states localized at the oxidized surface peroxo group, as well as other hole and electron states which are not associated with the surface peroxo group. 

This hybridized excitation nature is also present upon the inclusion of sub-surface defects in the oxidized structure, Fig.~\ref{excitons}c,d. The lowest bright excitation is at 3.08~eV, and the low-energy absorption peak is at 3.22~eV, similarly to the pristine case. In contrast to the non-defective structure, in this case, defect-induced localized oxo states play a significant role in the excitation. The electron-hole contribution from defect-induced oxide states are shown as black dots in Fig.~\ref{excitons}c and the associated wavefunctions are presented in Fig.~\ref{excitons}d. Although these defect-induced states show significant electron-hole coupling, our GW-BSE results suggest that the associated excitons strongly mix defect-induced transitions with non-defect transitions. As a result, the overall peak position does not change compared to the non-defect case. Since the quasiparticle gap is also similar in both systems, the exciton binding energy associated with the low-energy excitation peak remains unchanged upon defect formation. However, the absorption contribution at the various polarization directions changes due to the modification in the underlying structure, reflecting change in directionality of the wavefunction components, as discussed above.     

The main excitation peaks thus show a highly hybridized nature, which eventually leads to comparable excitation energies between the non-defect and the defect oxidized surface structures. 
This is a direct outcome of the localized nature of both non-defect and defect states in this system, as is also evident by the large quasiparticle gap and the electron-hole binding energies found with and without defects. This suggests that the role of oxygen vacancies in the associated catalytic cycle mainly comes to play in structural modification and stabilization upon water oxidation. Such modification leads to near-gap surface states which can support improved charge transfer, in particular upon defect charging expected to occur through electronic transport and push those bands deeper into the gap~\cite{seo2018role}. However, our computations show that the electron-hole transitions composing the excitons are hybridized and include both pristine and defect states, suggesting that defects cannot be trivially treated as charge traps. Our results thus imply that the many-body excitonic picture associated with photogeneration processes within BiVO$_4$ surfaces photocatalysis in the presence of defects can support improved carrier transport.

To conclude, in this work we explored how sub-surface defects alter the electronic and excitonic properties in BiVO$_4$ within atomic structures associated with surface water oxidation. We find that state localization due to structural modifications strongly influences the quasiparticle charge distributions and energy levels. However, oxygen vacancies only slightly change the exciton states, due to hybridized electron-hole transitions which mix defect and pristine states at similar energy regions. Our study presents a connection between the change in chemical bonding due to the presence of defects and the subsequent variations in the electronic bandstructure, wavefunctions, and optical excitations, demonstrating the non-trivial effect of defects on the mechanisms in which light is stored and energy is transferred in BiVO$_4$. 

Acknowledgements: We thank Diana Y. Qiu and Felipe H. da Jornada for helpful discussions. This research was supported by an Israel Science Foundation Grant No.1208/19. Computational resources were provided
by the Oak Ridge Leadership Computing Facility through the Innovative and Novel Computational Impact on Theory and Experiment (INCITE) program, which is a DOE Office of Science User Facility supported under Contract No. DE-AC05-00OR22725. Additional computational resources were provided by the ChemFarm local cluster at the Weizmann Institute of Science. R.S-E. and D.H.-P. acknowledge funding from a Minerva Foundation grant. S.R.A. is an incumbent of the Leah Omenn Career Development Chair and acknowledges a Peter and Patricia Gruber Award and an Alon Fellowship. 
%\end{acknowledgements}

%\bibliography{ref.bib}

\begin{thebibliography}{52}%
\makeatletter
\providecommand \@ifxundefined [1]{%
 \@ifx{#1\undefined}
}%
\providecommand \@ifnum [1]{%
 \ifnum #1\expandafter \@firstoftwo
 \else \expandafter \@secondoftwo
 \fi
}%
\providecommand \@ifx [1]{%
 \ifx #1\expandafter \@firstoftwo
 \else \expandafter \@secondoftwo
 \fi
}%
\providecommand \natexlab [1]{#1}%
\providecommand \enquote  [1]{``#1''}%
\providecommand \bibnamefont  [1]{#1}%
\providecommand \bibfnamefont [1]{#1}%
\providecommand \citenamefont [1]{#1}%
\providecommand \href@noop [0]{\@secondoftwo}%
\providecommand \href [0]{\begingroup \@sanitize@url \@href}%
\providecommand \@href[1]{\@@startlink{#1}\@@href}%
\providecommand \@@href[1]{\endgroup#1\@@endlink}%
\providecommand \@sanitize@url [0]{\catcode `\\12\catcode `\$12\catcode
  `\&12\catcode `\#12\catcode `\^12\catcode `\_12\catcode `\%12\relax}%
\providecommand \@@startlink[1]{}%
\providecommand \@@endlink[0]{}%
\providecommand \url  [0]{\begingroup\@sanitize@url \@url }%
\providecommand \@url [1]{\endgroup\@href {#1}{\urlprefix }}%
\providecommand \urlprefix  [0]{URL }%
\providecommand \Eprint [0]{\href }%
\providecommand \doibase [0]{http://dx.doi.org/}%
\providecommand \selectlanguage [0]{\@gobble}%
\providecommand \bibinfo  [0]{\@secondoftwo}%
\providecommand \bibfield  [0]{\@secondoftwo}%
\providecommand \translation [1]{[#1]}%
\providecommand \BibitemOpen [0]{}%
\providecommand \bibitemStop [0]{}%
\providecommand \bibitemNoStop [0]{.\EOS\space}%
\providecommand \EOS [0]{\spacefactor3000\relax}%
\providecommand \BibitemShut  [1]{\csname bibitem#1\endcsname}%
\let\auto@bib@innerbib\@empty
%</preamble>
\bibitem [{\citenamefont {Jiang}\ \emph {et~al.}(2017)\citenamefont {Jiang},
  \citenamefont {Moniz}, \citenamefont {Wang}, \citenamefont {Zhang},\ and\
  \citenamefont {Tang}}]{jiang2017photoelectrochemical}%
  \BibitemOpen
  \bibfield  {author} {\bibinfo {author} {\bibfnamefont {C.}~\bibnamefont
  {Jiang}}, \bibinfo {author} {\bibfnamefont {S.~J.}\ \bibnamefont {Moniz}},
  \bibinfo {author} {\bibfnamefont {A.}~\bibnamefont {Wang}}, \bibinfo {author}
  {\bibfnamefont {T.}~\bibnamefont {Zhang}}, \ and\ \bibinfo {author}
  {\bibfnamefont {J.}~\bibnamefont {Tang}},\ }\href@noop {} {\bibfield
  {journal} {\bibinfo  {journal} {Chemical Society Reviews}\ }\textbf {\bibinfo
  {volume} {46}},\ \bibinfo {pages} {4645} (\bibinfo {year}
  {2017})}\BibitemShut {NoStop}%
\bibitem [{\citenamefont {Jia}\ \emph {et~al.}(2012)\citenamefont {Jia},
  \citenamefont {Iwashina},\ and\ \citenamefont {Kudo}}]{jia2012facile}%
  \BibitemOpen
  \bibfield  {author} {\bibinfo {author} {\bibfnamefont {Q.}~\bibnamefont
  {Jia}}, \bibinfo {author} {\bibfnamefont {K.}~\bibnamefont {Iwashina}}, \
  and\ \bibinfo {author} {\bibfnamefont {A.}~\bibnamefont {Kudo}},\ }\href@noop
  {} {\bibfield  {journal} {\bibinfo  {journal} {Proceedings of the National
  Academy of Sciences}\ }\textbf {\bibinfo {volume} {109}},\ \bibinfo {pages}
  {11564} (\bibinfo {year} {2012})}\BibitemShut {NoStop}%
\bibitem [{\citenamefont {Kudo}\ \emph {et~al.}(1998)\citenamefont {Kudo},
  \citenamefont {Ueda}, \citenamefont {Kato},\ and\ \citenamefont
  {Mikami}}]{kudo1998photocatalytic}%
  \BibitemOpen
  \bibfield  {author} {\bibinfo {author} {\bibfnamefont {A.}~\bibnamefont
  {Kudo}}, \bibinfo {author} {\bibfnamefont {K.}~\bibnamefont {Ueda}}, \bibinfo
  {author} {\bibfnamefont {H.}~\bibnamefont {Kato}}, \ and\ \bibinfo {author}
  {\bibfnamefont {I.}~\bibnamefont {Mikami}},\ }\href@noop {} {\bibfield
  {journal} {\bibinfo  {journal} {Catalysis Letters}\ }\textbf {\bibinfo
  {volume} {53}},\ \bibinfo {pages} {229} (\bibinfo {year} {1998})}\BibitemShut
  {NoStop}%
\bibitem [{\citenamefont {Sharp}\ \emph {et~al.}(2017)\citenamefont {Sharp},
  \citenamefont {Cooper}, \citenamefont {Toma},\ and\ \citenamefont
  {Buonsanti}}]{sharp2017bismuth}%
  \BibitemOpen
  \bibfield  {author} {\bibinfo {author} {\bibfnamefont {I.~D.}\ \bibnamefont
  {Sharp}}, \bibinfo {author} {\bibfnamefont {J.~K.}\ \bibnamefont {Cooper}},
  \bibinfo {author} {\bibfnamefont {F.~M.}\ \bibnamefont {Toma}}, \ and\
  \bibinfo {author} {\bibfnamefont {R.}~\bibnamefont {Buonsanti}},\ }\href@noop
  {} {\bibfield  {journal} {\bibinfo  {journal} {ACS Energy Letters}\ }\textbf
  {\bibinfo {volume} {2}},\ \bibinfo {pages} {139} (\bibinfo {year}
  {2017})}\BibitemShut {NoStop}%
\bibitem [{\citenamefont {Tan}\ \emph {et~al.}(2017)\citenamefont {Tan},
  \citenamefont {Amal},\ and\ \citenamefont {Ng}}]{tan2017alternative}%
  \BibitemOpen
  \bibfield  {author} {\bibinfo {author} {\bibfnamefont {H.~L.}\ \bibnamefont
  {Tan}}, \bibinfo {author} {\bibfnamefont {R.}~\bibnamefont {Amal}}, \ and\
  \bibinfo {author} {\bibfnamefont {Y.~H.}\ \bibnamefont {Ng}},\ }\href@noop {}
  {\bibfield  {journal} {\bibinfo  {journal} {Journal of Materials Chemistry
  A}\ }\textbf {\bibinfo {volume} {5}},\ \bibinfo {pages} {16498} (\bibinfo
  {year} {2017})}\BibitemShut {NoStop}%
\bibitem [{\citenamefont {Tolod}\ \emph {et~al.}(2017)\citenamefont {Tolod},
  \citenamefont {Hern{\'a}ndez},\ and\ \citenamefont
  {Russo}}]{tolod2017recent}%
  \BibitemOpen
  \bibfield  {author} {\bibinfo {author} {\bibfnamefont {K.~R.}\ \bibnamefont
  {Tolod}}, \bibinfo {author} {\bibfnamefont {S.}~\bibnamefont
  {Hern{\'a}ndez}}, \ and\ \bibinfo {author} {\bibfnamefont {N.}~\bibnamefont
  {Russo}},\ }\href@noop {} {\bibfield  {journal} {\bibinfo  {journal}
  {Catalysts}\ }\textbf {\bibinfo {volume} {7}},\ \bibinfo {pages} {13}
  (\bibinfo {year} {2017})}\BibitemShut {NoStop}%
\bibitem [{\citenamefont {Zhao}\ \emph {et~al.}(2011)\citenamefont {Zhao},
  \citenamefont {Li},\ and\ \citenamefont {Zou}}]{zhao2011electronic}%
  \BibitemOpen
  \bibfield  {author} {\bibinfo {author} {\bibfnamefont {Z.}~\bibnamefont
  {Zhao}}, \bibinfo {author} {\bibfnamefont {Z.}~\bibnamefont {Li}}, \ and\
  \bibinfo {author} {\bibfnamefont {Z.}~\bibnamefont {Zou}},\ }\href@noop {}
  {\bibfield  {journal} {\bibinfo  {journal} {Physical Chemistry Chemical
  Physics}\ }\textbf {\bibinfo {volume} {13}},\ \bibinfo {pages} {4746}
  (\bibinfo {year} {2011})}\BibitemShut {NoStop}%
\bibitem [{\citenamefont {Wiktor}\ \emph {et~al.}(2017)\citenamefont {Wiktor},
  \citenamefont {Reshetnyak}, \citenamefont {Ambrosio},\ and\ \citenamefont
  {Pasquarello}}]{wiktor2017comprehensive}%
  \BibitemOpen
  \bibfield  {author} {\bibinfo {author} {\bibfnamefont {J.}~\bibnamefont
  {Wiktor}}, \bibinfo {author} {\bibfnamefont {I.}~\bibnamefont {Reshetnyak}},
  \bibinfo {author} {\bibfnamefont {F.}~\bibnamefont {Ambrosio}}, \ and\
  \bibinfo {author} {\bibfnamefont {A.}~\bibnamefont {Pasquarello}},\
  }\href@noop {} {\bibfield  {journal} {\bibinfo  {journal} {Physical Review
  Materials}\ }\textbf {\bibinfo {volume} {1}},\ \bibinfo {pages} {022401}
  (\bibinfo {year} {2017})}\BibitemShut {NoStop}%
\bibitem [{\citenamefont {Das}\ \emph {et~al.}(2017)\citenamefont {Das},
  \citenamefont {Rocquefelte}, \citenamefont {Laskowski}, \citenamefont
  {Lajaunie}, \citenamefont {Jobic}, \citenamefont {Blaha},\ and\ \citenamefont
  {Schwarz}}]{das2017investigation}%
  \BibitemOpen
  \bibfield  {author} {\bibinfo {author} {\bibfnamefont {T.}~\bibnamefont
  {Das}}, \bibinfo {author} {\bibfnamefont {X.}~\bibnamefont {Rocquefelte}},
  \bibinfo {author} {\bibfnamefont {R.}~\bibnamefont {Laskowski}}, \bibinfo
  {author} {\bibfnamefont {L.}~\bibnamefont {Lajaunie}}, \bibinfo {author}
  {\bibfnamefont {S.}~\bibnamefont {Jobic}}, \bibinfo {author} {\bibfnamefont
  {P.}~\bibnamefont {Blaha}}, \ and\ \bibinfo {author} {\bibfnamefont
  {K.}~\bibnamefont {Schwarz}},\ }\href@noop {} {\bibfield  {journal} {\bibinfo
   {journal} {Chemistry of Materials}\ }\textbf {\bibinfo {volume} {29}},\
  \bibinfo {pages} {3380} (\bibinfo {year} {2017})}\BibitemShut {NoStop}%
\bibitem [{\citenamefont {Wang}\ \emph {et~al.}(2021)\citenamefont {Wang},
  \citenamefont {Shi}, \citenamefont {Ge}, \citenamefont {Zhang}, \citenamefont
  {Wang},\ and\ \citenamefont {Yu}}]{wang2021exciton}%
  \BibitemOpen
  \bibfield  {author} {\bibinfo {author} {\bibfnamefont {Y.}~\bibnamefont
  {Wang}}, \bibinfo {author} {\bibfnamefont {H.}~\bibnamefont {Shi}}, \bibinfo
  {author} {\bibfnamefont {S.}~\bibnamefont {Ge}}, \bibinfo {author}
  {\bibfnamefont {L.}~\bibnamefont {Zhang}}, \bibinfo {author} {\bibfnamefont
  {X.}~\bibnamefont {Wang}}, \ and\ \bibinfo {author} {\bibfnamefont
  {J.}~\bibnamefont {Yu}},\ }\href@noop {} {\bibfield  {journal} {\bibinfo
  {journal} {Sensors and Actuators B: Chemical}\ }\textbf {\bibinfo {volume}
  {336}},\ \bibinfo {pages} {129746} (\bibinfo {year} {2021})}\BibitemShut
  {NoStop}%
\bibitem [{\citenamefont {Cooper}\ \emph {et~al.}(2014)\citenamefont {Cooper},
  \citenamefont {Gul}, \citenamefont {Toma}, \citenamefont {Chen},
  \citenamefont {Glans}, \citenamefont {Guo}, \citenamefont {Ager},
  \citenamefont {Yano},\ and\ \citenamefont {Sharp}}]{cooper2014electronic}%
  \BibitemOpen
  \bibfield  {author} {\bibinfo {author} {\bibfnamefont {J.~K.}\ \bibnamefont
  {Cooper}}, \bibinfo {author} {\bibfnamefont {S.}~\bibnamefont {Gul}},
  \bibinfo {author} {\bibfnamefont {F.~M.}\ \bibnamefont {Toma}}, \bibinfo
  {author} {\bibfnamefont {L.}~\bibnamefont {Chen}}, \bibinfo {author}
  {\bibfnamefont {P.-A.}\ \bibnamefont {Glans}}, \bibinfo {author}
  {\bibfnamefont {J.}~\bibnamefont {Guo}}, \bibinfo {author} {\bibfnamefont
  {J.~W.}\ \bibnamefont {Ager}}, \bibinfo {author} {\bibfnamefont
  {J.}~\bibnamefont {Yano}}, \ and\ \bibinfo {author} {\bibfnamefont {I.~D.}\
  \bibnamefont {Sharp}},\ }\href@noop {} {\bibfield  {journal} {\bibinfo
  {journal} {Chemistry of Materials}\ }\textbf {\bibinfo {volume} {26}},\
  \bibinfo {pages} {5365} (\bibinfo {year} {2014})}\BibitemShut {NoStop}%
\bibitem [{\citenamefont {Park}\ \emph {et~al.}(2013)\citenamefont {Park},
  \citenamefont {McDonald},\ and\ \citenamefont {Choi}}]{park2013progress}%
  \BibitemOpen
  \bibfield  {author} {\bibinfo {author} {\bibfnamefont {Y.}~\bibnamefont
  {Park}}, \bibinfo {author} {\bibfnamefont {K.~J.}\ \bibnamefont {McDonald}},
  \ and\ \bibinfo {author} {\bibfnamefont {K.-S.}\ \bibnamefont {Choi}},\
  }\href@noop {} {\bibfield  {journal} {\bibinfo  {journal} {Chemical Society
  Reviews}\ }\textbf {\bibinfo {volume} {42}},\ \bibinfo {pages} {2321}
  (\bibinfo {year} {2013})}\BibitemShut {NoStop}%
\bibitem [{\citenamefont {Abdi}\ \emph
  {et~al.}(2013{\natexlab{a}})\citenamefont {Abdi}, \citenamefont {Savenije},
  \citenamefont {May}, \citenamefont {Dam},\ and\ \citenamefont {van~de
  Krol}}]{abdi2013origin}%
  \BibitemOpen
  \bibfield  {author} {\bibinfo {author} {\bibfnamefont {F.~F.}\ \bibnamefont
  {Abdi}}, \bibinfo {author} {\bibfnamefont {T.~J.}\ \bibnamefont {Savenije}},
  \bibinfo {author} {\bibfnamefont {M.~M.}\ \bibnamefont {May}}, \bibinfo
  {author} {\bibfnamefont {B.}~\bibnamefont {Dam}}, \ and\ \bibinfo {author}
  {\bibfnamefont {R.}~\bibnamefont {van~de Krol}},\ }\href@noop {} {\bibfield
  {journal} {\bibinfo  {journal} {The Journal of Physical Chemistry Letters}\
  }\textbf {\bibinfo {volume} {4}},\ \bibinfo {pages} {2752} (\bibinfo {year}
  {2013}{\natexlab{a}})}\BibitemShut {NoStop}%
\bibitem [{\citenamefont {Rettie}\ \emph {et~al.}(2013)\citenamefont {Rettie},
  \citenamefont {Lee}, \citenamefont {Marshall}, \citenamefont {Lin},
  \citenamefont {Capan}, \citenamefont {Lindemuth}, \citenamefont {McCloy},
  \citenamefont {Zhou}, \citenamefont {Bard},\ and\ \citenamefont
  {Mullins}}]{rettie2013combined}%
  \BibitemOpen
  \bibfield  {author} {\bibinfo {author} {\bibfnamefont {A.~J.}\ \bibnamefont
  {Rettie}}, \bibinfo {author} {\bibfnamefont {H.~C.}\ \bibnamefont {Lee}},
  \bibinfo {author} {\bibfnamefont {L.~G.}\ \bibnamefont {Marshall}}, \bibinfo
  {author} {\bibfnamefont {J.-F.}\ \bibnamefont {Lin}}, \bibinfo {author}
  {\bibfnamefont {C.}~\bibnamefont {Capan}}, \bibinfo {author} {\bibfnamefont
  {J.}~\bibnamefont {Lindemuth}}, \bibinfo {author} {\bibfnamefont {J.~S.}\
  \bibnamefont {McCloy}}, \bibinfo {author} {\bibfnamefont {J.}~\bibnamefont
  {Zhou}}, \bibinfo {author} {\bibfnamefont {A.~J.}\ \bibnamefont {Bard}}, \
  and\ \bibinfo {author} {\bibfnamefont {C.~B.}\ \bibnamefont {Mullins}},\
  }\href@noop {} {\bibfield  {journal} {\bibinfo  {journal} {Journal of the
  American Chemical Society}\ }\textbf {\bibinfo {volume} {135}},\ \bibinfo
  {pages} {11389} (\bibinfo {year} {2013})}\BibitemShut {NoStop}%
\bibitem [{\citenamefont {Seo}\ \emph {et~al.}(2018)\citenamefont {Seo},
  \citenamefont {Ping},\ and\ \citenamefont {Galli}}]{seo2018role}%
  \BibitemOpen
  \bibfield  {author} {\bibinfo {author} {\bibfnamefont {H.}~\bibnamefont
  {Seo}}, \bibinfo {author} {\bibfnamefont {Y.}~\bibnamefont {Ping}}, \ and\
  \bibinfo {author} {\bibfnamefont {G.}~\bibnamefont {Galli}},\ }\href@noop {}
  {\bibfield  {journal} {\bibinfo  {journal} {Chemistry of Materials}\ }\textbf
  {\bibinfo {volume} {30}},\ \bibinfo {pages} {7793} (\bibinfo {year}
  {2018})}\BibitemShut {NoStop}%
\bibitem [{\citenamefont {Kim}\ \emph {et~al.}(2015)\citenamefont {Kim},
  \citenamefont {Ping}, \citenamefont {Galli},\ and\ \citenamefont
  {Choi}}]{kim2015simultaneous}%
  \BibitemOpen
  \bibfield  {author} {\bibinfo {author} {\bibfnamefont {T.~W.}\ \bibnamefont
  {Kim}}, \bibinfo {author} {\bibfnamefont {Y.}~\bibnamefont {Ping}}, \bibinfo
  {author} {\bibfnamefont {G.~A.}\ \bibnamefont {Galli}}, \ and\ \bibinfo
  {author} {\bibfnamefont {K.-S.}\ \bibnamefont {Choi}},\ }\href@noop {}
  {\bibfield  {journal} {\bibinfo  {journal} {Nature Communications}\ }\textbf
  {\bibinfo {volume} {6}},\ \bibinfo {pages} {1} (\bibinfo {year}
  {2015})}\BibitemShut {NoStop}%
\bibitem [{\citenamefont {Park}\ \emph {et~al.}(2011)\citenamefont {Park},
  \citenamefont {Kweon}, \citenamefont {Ye}, \citenamefont {Paek},
  \citenamefont {Hwang},\ and\ \citenamefont {Bard}}]{park2011factors}%
  \BibitemOpen
  \bibfield  {author} {\bibinfo {author} {\bibfnamefont {H.~S.}\ \bibnamefont
  {Park}}, \bibinfo {author} {\bibfnamefont {K.~E.}\ \bibnamefont {Kweon}},
  \bibinfo {author} {\bibfnamefont {H.}~\bibnamefont {Ye}}, \bibinfo {author}
  {\bibfnamefont {E.}~\bibnamefont {Paek}}, \bibinfo {author} {\bibfnamefont
  {G.~S.}\ \bibnamefont {Hwang}}, \ and\ \bibinfo {author} {\bibfnamefont
  {A.~J.}\ \bibnamefont {Bard}},\ }\href@noop {} {\bibfield  {journal}
  {\bibinfo  {journal} {The Journal of Physical Chemistry C}\ }\textbf
  {\bibinfo {volume} {115}},\ \bibinfo {pages} {17870} (\bibinfo {year}
  {2011})}\BibitemShut {NoStop}%
\bibitem [{\citenamefont {Wang}\ \emph {et~al.}(2018)\citenamefont {Wang},
  \citenamefont {Chen}, \citenamefont {Bai}, \citenamefont {Yun}, \citenamefont
  {Liu},\ and\ \citenamefont {Wang}}]{wang2018new}%
  \BibitemOpen
  \bibfield  {author} {\bibinfo {author} {\bibfnamefont {S.}~\bibnamefont
  {Wang}}, \bibinfo {author} {\bibfnamefont {P.}~\bibnamefont {Chen}}, \bibinfo
  {author} {\bibfnamefont {Y.}~\bibnamefont {Bai}}, \bibinfo {author}
  {\bibfnamefont {J.-H.}\ \bibnamefont {Yun}}, \bibinfo {author} {\bibfnamefont
  {G.}~\bibnamefont {Liu}}, \ and\ \bibinfo {author} {\bibfnamefont
  {L.}~\bibnamefont {Wang}},\ }\href@noop {} {\bibfield  {journal} {\bibinfo
  {journal} {Advanced Materials}\ }\textbf {\bibinfo {volume} {30}},\ \bibinfo
  {pages} {1800486} (\bibinfo {year} {2018})}\BibitemShut {NoStop}%
\bibitem [{\citenamefont {Hammes-Schiffer}\ and\ \citenamefont
  {Galli}(2021)}]{hammes2021integration}%
  \BibitemOpen
  \bibfield  {author} {\bibinfo {author} {\bibfnamefont {S.}~\bibnamefont
  {Hammes-Schiffer}}\ and\ \bibinfo {author} {\bibfnamefont {G.}~\bibnamefont
  {Galli}},\ }\href {\doibase 10.1038/s41560-021-00827-4} {\bibfield  {journal}
  {\bibinfo  {journal} {Nature Energy}\ }\textbf {\bibinfo {volume} {6}},\
  \bibinfo {pages} {700} (\bibinfo {year} {2021})}\BibitemShut {NoStop}%
\bibitem [{\citenamefont {Hu}\ \emph {et~al.}(2018)\citenamefont {Hu},
  \citenamefont {Chen}, \citenamefont {Zhao}, \citenamefont {Su},\ and\
  \citenamefont {Chen}}]{hu2018anisotropic}%
  \BibitemOpen
  \bibfield  {author} {\bibinfo {author} {\bibfnamefont {J.}~\bibnamefont
  {Hu}}, \bibinfo {author} {\bibfnamefont {W.}~\bibnamefont {Chen}}, \bibinfo
  {author} {\bibfnamefont {X.}~\bibnamefont {Zhao}}, \bibinfo {author}
  {\bibfnamefont {H.}~\bibnamefont {Su}}, \ and\ \bibinfo {author}
  {\bibfnamefont {Z.}~\bibnamefont {Chen}},\ }\href@noop {} {\bibfield
  {journal} {\bibinfo  {journal} {ACS Applied Materials \& Interfaces}\
  }\textbf {\bibinfo {volume} {10}},\ \bibinfo {pages} {5475} (\bibinfo {year}
  {2018})}\BibitemShut {NoStop}%
\bibitem [{\citenamefont {Li}\ \emph {et~al.}(2019)\citenamefont {Li},
  \citenamefont {Liu}, \citenamefont {Shi}, \citenamefont {Shao}, \citenamefont
  {Wang}, \citenamefont {Ding}, \citenamefont {Liu}, \citenamefont {Fan},
  \citenamefont {Shi},\ and\ \citenamefont {Li}}]{li2019crystallographic}%
  \BibitemOpen
  \bibfield  {author} {\bibinfo {author} {\bibfnamefont {D.}~\bibnamefont
  {Li}}, \bibinfo {author} {\bibfnamefont {Y.}~\bibnamefont {Liu}}, \bibinfo
  {author} {\bibfnamefont {W.}~\bibnamefont {Shi}}, \bibinfo {author}
  {\bibfnamefont {C.}~\bibnamefont {Shao}}, \bibinfo {author} {\bibfnamefont
  {S.}~\bibnamefont {Wang}}, \bibinfo {author} {\bibfnamefont {C.}~\bibnamefont
  {Ding}}, \bibinfo {author} {\bibfnamefont {T.}~\bibnamefont {Liu}}, \bibinfo
  {author} {\bibfnamefont {F.}~\bibnamefont {Fan}}, \bibinfo {author}
  {\bibfnamefont {J.}~\bibnamefont {Shi}}, \ and\ \bibinfo {author}
  {\bibfnamefont {C.}~\bibnamefont {Li}},\ }\href@noop {} {\bibfield  {journal}
  {\bibinfo  {journal} {ACS Energy Letters}\ }\textbf {\bibinfo {volume} {4}},\
  \bibinfo {pages} {825} (\bibinfo {year} {2019})}\BibitemShut {NoStop}%
\bibitem [{\citenamefont {Wu}\ \emph {et~al.}(2021)\citenamefont {Wu},
  \citenamefont {Irani}, \citenamefont {Zhang}, \citenamefont {Jing},
  \citenamefont {Dai}, \citenamefont {Chung}, \citenamefont {Abdi},\ and\
  \citenamefont {Ng}}]{wu2021unveiling}%
  \BibitemOpen
  \bibfield  {author} {\bibinfo {author} {\bibfnamefont {H.}~\bibnamefont
  {Wu}}, \bibinfo {author} {\bibfnamefont {R.}~\bibnamefont {Irani}}, \bibinfo
  {author} {\bibfnamefont {K.}~\bibnamefont {Zhang}}, \bibinfo {author}
  {\bibfnamefont {L.}~\bibnamefont {Jing}}, \bibinfo {author} {\bibfnamefont
  {H.}~\bibnamefont {Dai}}, \bibinfo {author} {\bibfnamefont {H.~Y.}\
  \bibnamefont {Chung}}, \bibinfo {author} {\bibfnamefont {F.~F.}\ \bibnamefont
  {Abdi}}, \ and\ \bibinfo {author} {\bibfnamefont {Y.~H.}\ \bibnamefont
  {Ng}},\ }\href@noop {} {\bibfield  {journal} {\bibinfo  {journal} {ACS Energy
  Letters}\ }\textbf {\bibinfo {volume} {6}},\ \bibinfo {pages} {3400}
  (\bibinfo {year} {2021})}\BibitemShut {NoStop}%
\bibitem [{\citenamefont {Lardhi}\ \emph {et~al.}(2020)\citenamefont {Lardhi},
  \citenamefont {Cavallo},\ and\ \citenamefont {Harb}}]{lardhi2020significant}%
  \BibitemOpen
  \bibfield  {author} {\bibinfo {author} {\bibfnamefont {S.}~\bibnamefont
  {Lardhi}}, \bibinfo {author} {\bibfnamefont {L.}~\bibnamefont {Cavallo}}, \
  and\ \bibinfo {author} {\bibfnamefont {M.}~\bibnamefont {Harb}},\ }\href@noop
  {} {\bibfield  {journal} {\bibinfo  {journal} {The Journal of Physical
  Chemistry Letters}\ }\textbf {\bibinfo {volume} {11}},\ \bibinfo {pages}
  {5497} (\bibinfo {year} {2020})}\BibitemShut {NoStop}%
\bibitem [{\citenamefont {Kahraman}\ \emph {et~al.}(2020)\citenamefont
  {Kahraman}, \citenamefont {Vishlaghi}, \citenamefont {Baylam}, \citenamefont
  {Ogasawara}, \citenamefont {Sennaroglu},\ and\ \citenamefont
  {Kaya}}]{kahraman2020fast}%
  \BibitemOpen
  \bibfield  {author} {\bibinfo {author} {\bibfnamefont {A.}~\bibnamefont
  {Kahraman}}, \bibinfo {author} {\bibfnamefont {M.~B.}\ \bibnamefont
  {Vishlaghi}}, \bibinfo {author} {\bibfnamefont {I.}~\bibnamefont {Baylam}},
  \bibinfo {author} {\bibfnamefont {H.}~\bibnamefont {Ogasawara}}, \bibinfo
  {author} {\bibfnamefont {A.}~\bibnamefont {Sennaroglu}}, \ and\ \bibinfo
  {author} {\bibfnamefont {S.}~\bibnamefont {Kaya}},\ }\href@noop {} {\bibfield
   {journal} {\bibinfo  {journal} {The Journal of Physical Chemistry Letters}\
  }\textbf {\bibinfo {volume} {11}},\ \bibinfo {pages} {8758} (\bibinfo {year}
  {2020})}\BibitemShut {NoStop}%
\bibitem [{\citenamefont {Nikac\v{e}vi{\'c}}\ \emph {et~al.}(2021)\citenamefont
  {Nikac\v{e}vi{\'c}}, \citenamefont {Hegner}, \citenamefont
  {Gal{\'a}n-Mascar{\'o}s},\ and\ \citenamefont {L\'opez}}]{pavle}%
  \BibitemOpen
  \bibfield  {author} {\bibinfo {author} {\bibfnamefont {P.}~\bibnamefont
  {Nikac\v{e}vi{\'c}}}, \bibinfo {author} {\bibfnamefont {F.~S.}\ \bibnamefont
  {Hegner}}, \bibinfo {author} {\bibfnamefont {J.~R.}\ \bibnamefont
  {Gal{\'a}n-Mascar{\'o}s}}, \ and\ \bibinfo {author} {\bibfnamefont
  {N.}~\bibnamefont {L\'opez}},\ }\href@noop {} {\bibfield  {journal} {\bibinfo
   {journal} {ACS Catalysis}\ }\textbf {\bibinfo {volume} {11}},\ \bibinfo
  {pages} {13416} (\bibinfo {year} {2021})}\BibitemShut {NoStop}%
\bibitem [{\citenamefont {Li}(2017)}]{li2017first}%
  \BibitemOpen
  \bibfield  {author} {\bibinfo {author} {\bibfnamefont {G.-L.}\ \bibnamefont
  {Li}},\ }\href@noop {} {\bibfield  {journal} {\bibinfo  {journal} {RSC
  Advances}\ }\textbf {\bibinfo {volume} {7}},\ \bibinfo {pages} {9130}
  (\bibinfo {year} {2017})}\BibitemShut {NoStop}%
\bibitem [{\citenamefont {Wang}\ \emph {et~al.}(2020)\citenamefont {Wang},
  \citenamefont {Strohbeen}, \citenamefont {Lee}, \citenamefont {Zhou},
  \citenamefont {Kawasaki}, \citenamefont {Choi}, \citenamefont {Liu},\ and\
  \citenamefont {Galli}}]{wang2020role}%
  \BibitemOpen
  \bibfield  {author} {\bibinfo {author} {\bibfnamefont {W.}~\bibnamefont
  {Wang}}, \bibinfo {author} {\bibfnamefont {P.~J.}\ \bibnamefont {Strohbeen}},
  \bibinfo {author} {\bibfnamefont {D.}~\bibnamefont {Lee}}, \bibinfo {author}
  {\bibfnamefont {C.}~\bibnamefont {Zhou}}, \bibinfo {author} {\bibfnamefont
  {J.~K.}\ \bibnamefont {Kawasaki}}, \bibinfo {author} {\bibfnamefont {K.-S.}\
  \bibnamefont {Choi}}, \bibinfo {author} {\bibfnamefont {M.}~\bibnamefont
  {Liu}}, \ and\ \bibinfo {author} {\bibfnamefont {G.}~\bibnamefont {Galli}},\
  }\href@noop {} {\bibfield  {journal} {\bibinfo  {journal} {Chemistry of
  Materials}\ }\textbf {\bibinfo {volume} {32}},\ \bibinfo {pages} {2899}
  (\bibinfo {year} {2020})}\BibitemShut {NoStop}%
\bibitem [{\citenamefont {Li}\ \emph {et~al.}(2015)\citenamefont {Li},
  \citenamefont {Wang}, \citenamefont {Jiang}, \citenamefont {Zheng},\ and\
  \citenamefont {Li}}]{li2015surfactant}%
  \BibitemOpen
  \bibfield  {author} {\bibinfo {author} {\bibfnamefont {D.}~\bibnamefont
  {Li}}, \bibinfo {author} {\bibfnamefont {W.}~\bibnamefont {Wang}}, \bibinfo
  {author} {\bibfnamefont {D.}~\bibnamefont {Jiang}}, \bibinfo {author}
  {\bibfnamefont {Y.}~\bibnamefont {Zheng}}, \ and\ \bibinfo {author}
  {\bibfnamefont {X.}~\bibnamefont {Li}},\ }\href@noop {} {\bibfield  {journal}
  {\bibinfo  {journal} {RSC Advances}\ }\textbf {\bibinfo {volume} {5}},\
  \bibinfo {pages} {14374} (\bibinfo {year} {2015})}\BibitemShut {NoStop}%
\bibitem [{\citenamefont {Fern{\'a}ndez-Climent}\ \emph
  {et~al.}(2020)\citenamefont {Fern{\'a}ndez-Climent}, \citenamefont
  {Gim{\'e}nez},\ and\ \citenamefont
  {Garc{\'\i}a-Tecedor}}]{fernandez2020role}%
  \BibitemOpen
  \bibfield  {author} {\bibinfo {author} {\bibfnamefont {R.}~\bibnamefont
  {Fern{\'a}ndez-Climent}}, \bibinfo {author} {\bibfnamefont {S.}~\bibnamefont
  {Gim{\'e}nez}}, \ and\ \bibinfo {author} {\bibfnamefont {M.}~\bibnamefont
  {Garc{\'\i}a-Tecedor}},\ }\href@noop {} {\bibfield  {journal} {\bibinfo
  {journal} {Sustainable Energy \& Fuels}\ }\textbf {\bibinfo {volume} {4}},\
  \bibinfo {pages} {5916} (\bibinfo {year} {2020})}\BibitemShut {NoStop}%
\bibitem [{\citenamefont {Luo}\ \emph {et~al.}(2011)\citenamefont {Luo},
  \citenamefont {Yang}, \citenamefont {Li}, \citenamefont {Zhang},
  \citenamefont {Liu}, \citenamefont {Zhao}, \citenamefont {Wang},
  \citenamefont {Yan}, \citenamefont {Yu},\ and\ \citenamefont
  {Zou}}]{luo2011solar}%
  \BibitemOpen
  \bibfield  {author} {\bibinfo {author} {\bibfnamefont {W.}~\bibnamefont
  {Luo}}, \bibinfo {author} {\bibfnamefont {Z.}~\bibnamefont {Yang}}, \bibinfo
  {author} {\bibfnamefont {Z.}~\bibnamefont {Li}}, \bibinfo {author}
  {\bibfnamefont {J.}~\bibnamefont {Zhang}}, \bibinfo {author} {\bibfnamefont
  {J.}~\bibnamefont {Liu}}, \bibinfo {author} {\bibfnamefont {Z.}~\bibnamefont
  {Zhao}}, \bibinfo {author} {\bibfnamefont {Z.}~\bibnamefont {Wang}}, \bibinfo
  {author} {\bibfnamefont {S.}~\bibnamefont {Yan}}, \bibinfo {author}
  {\bibfnamefont {T.}~\bibnamefont {Yu}}, \ and\ \bibinfo {author}
  {\bibfnamefont {Z.}~\bibnamefont {Zou}},\ }\href@noop {} {\bibfield
  {journal} {\bibinfo  {journal} {Energy \& Environmental Science}\ }\textbf
  {\bibinfo {volume} {4}},\ \bibinfo {pages} {4046} (\bibinfo {year}
  {2011})}\BibitemShut {NoStop}%
\bibitem [{\citenamefont {Abdi}\ \emph
  {et~al.}(2013{\natexlab{b}})\citenamefont {Abdi}, \citenamefont {Han},
  \citenamefont {Smets}, \citenamefont {Zeman}, \citenamefont {Dam},\ and\
  \citenamefont {Van De~Krol}}]{abdi2013efficient}%
  \BibitemOpen
  \bibfield  {author} {\bibinfo {author} {\bibfnamefont {F.~F.}\ \bibnamefont
  {Abdi}}, \bibinfo {author} {\bibfnamefont {L.}~\bibnamefont {Han}}, \bibinfo
  {author} {\bibfnamefont {A.~H.}\ \bibnamefont {Smets}}, \bibinfo {author}
  {\bibfnamefont {M.}~\bibnamefont {Zeman}}, \bibinfo {author} {\bibfnamefont
  {B.}~\bibnamefont {Dam}}, \ and\ \bibinfo {author} {\bibfnamefont
  {R.}~\bibnamefont {Van De~Krol}},\ }\href@noop {} {\bibfield  {journal}
  {\bibinfo  {journal} {Nature communications}\ }\textbf {\bibinfo {volume}
  {4}},\ \bibinfo {pages} {1} (\bibinfo {year}
  {2013}{\natexlab{b}})}\BibitemShut {NoStop}%
\bibitem [{\citenamefont {Liang}\ \emph {et~al.}(2011)\citenamefont {Liang},
  \citenamefont {Tsubota}, \citenamefont {Mooij},\ and\ \citenamefont {van~de
  Krol}}]{liang2011highly}%
  \BibitemOpen
  \bibfield  {author} {\bibinfo {author} {\bibfnamefont {Y.}~\bibnamefont
  {Liang}}, \bibinfo {author} {\bibfnamefont {T.}~\bibnamefont {Tsubota}},
  \bibinfo {author} {\bibfnamefont {L.~P.}\ \bibnamefont {Mooij}}, \ and\
  \bibinfo {author} {\bibfnamefont {R.}~\bibnamefont {van~de Krol}},\
  }\href@noop {} {\bibfield  {journal} {\bibinfo  {journal} {The Journal of
  Physical Chemistry C}\ }\textbf {\bibinfo {volume} {115}},\ \bibinfo {pages}
  {17594} (\bibinfo {year} {2011})}\BibitemShut {NoStop}%
\bibitem [{\citenamefont {Cooper}\ \emph {et~al.}(2016)\citenamefont {Cooper},
  \citenamefont {Scott}, \citenamefont {Ling}, \citenamefont {Yang},
  \citenamefont {Hao}, \citenamefont {Li}, \citenamefont {Toma}, \citenamefont
  {Stutzmann}, \citenamefont {Lakshmi},\ and\ \citenamefont
  {Sharp}}]{cooper2016role}%
  \BibitemOpen
  \bibfield  {author} {\bibinfo {author} {\bibfnamefont {J.~K.}\ \bibnamefont
  {Cooper}}, \bibinfo {author} {\bibfnamefont {S.~B.}\ \bibnamefont {Scott}},
  \bibinfo {author} {\bibfnamefont {Y.}~\bibnamefont {Ling}}, \bibinfo {author}
  {\bibfnamefont {J.}~\bibnamefont {Yang}}, \bibinfo {author} {\bibfnamefont
  {S.}~\bibnamefont {Hao}}, \bibinfo {author} {\bibfnamefont {Y.}~\bibnamefont
  {Li}}, \bibinfo {author} {\bibfnamefont {F.~M.}\ \bibnamefont {Toma}},
  \bibinfo {author} {\bibfnamefont {M.}~\bibnamefont {Stutzmann}}, \bibinfo
  {author} {\bibfnamefont {K.}~\bibnamefont {Lakshmi}}, \ and\ \bibinfo
  {author} {\bibfnamefont {I.~D.}\ \bibnamefont {Sharp}},\ }\href@noop {}
  {\bibfield  {journal} {\bibinfo  {journal} {Chemistry of Materials}\ }\textbf
  {\bibinfo {volume} {28}},\ \bibinfo {pages} {5761} (\bibinfo {year}
  {2016})}\BibitemShut {NoStop}%
\bibitem [{\citenamefont {Daelman}\ \emph {et~al.}(2020)\citenamefont
  {Daelman}, \citenamefont {Hegner}, \citenamefont {Rell{\'a}n-Pi{\~n}eiro},
  \citenamefont {Capdevila-Cortada}, \citenamefont {Garc{\'\i}a-Muelas},\ and\
  \citenamefont {L{\'o}pez}}]{daelman2020quasi}%
  \BibitemOpen
  \bibfield  {author} {\bibinfo {author} {\bibfnamefont {N.}~\bibnamefont
  {Daelman}}, \bibinfo {author} {\bibfnamefont {F.~S.}\ \bibnamefont {Hegner}},
  \bibinfo {author} {\bibfnamefont {M.}~\bibnamefont {Rell{\'a}n-Pi{\~n}eiro}},
  \bibinfo {author} {\bibfnamefont {M.}~\bibnamefont {Capdevila-Cortada}},
  \bibinfo {author} {\bibfnamefont {R.}~\bibnamefont {Garc{\'\i}a-Muelas}}, \
  and\ \bibinfo {author} {\bibfnamefont {N.}~\bibnamefont {L{\'o}pez}},\
  }\href@noop {} {\bibfield  {journal} {\bibinfo  {journal} {The Journal of
  Chemical Physics}\ }\textbf {\bibinfo {volume} {152}},\ \bibinfo {pages}
  {050901} (\bibinfo {year} {2020})}\BibitemShut {NoStop}%
\bibitem [{\citenamefont {Hegner}\ \emph {et~al.}(2019)\citenamefont {Hegner},
  \citenamefont {Forrer}, \citenamefont {Galan-Mascaros}, \citenamefont
  {L{\'o}pez},\ and\ \citenamefont {Selloni}}]{hegner2019versatile}%
  \BibitemOpen
  \bibfield  {author} {\bibinfo {author} {\bibfnamefont {F.~S.}\ \bibnamefont
  {Hegner}}, \bibinfo {author} {\bibfnamefont {D.}~\bibnamefont {Forrer}},
  \bibinfo {author} {\bibfnamefont {J.~R.}\ \bibnamefont {Galan-Mascaros}},
  \bibinfo {author} {\bibfnamefont {N.}~\bibnamefont {L{\'o}pez}}, \ and\
  \bibinfo {author} {\bibfnamefont {A.}~\bibnamefont {Selloni}},\ }\href@noop
  {} {\bibfield  {journal} {\bibinfo  {journal} {The Journal of Physical
  Chemistry Letters}\ }\textbf {\bibinfo {volume} {10}},\ \bibinfo {pages}
  {6672} (\bibinfo {year} {2019})}\BibitemShut {NoStop}%
\bibitem [{\citenamefont {Ulpe}\ \emph {et~al.}(2018)\citenamefont {Ulpe},
  \citenamefont {Anke}, \citenamefont {Berendts}, \citenamefont {Lerch},\ and\
  \citenamefont {Bredow}}]{ulpe2018f}%
  \BibitemOpen
  \bibfield  {author} {\bibinfo {author} {\bibfnamefont {A.~C.}\ \bibnamefont
  {Ulpe}}, \bibinfo {author} {\bibfnamefont {B.}~\bibnamefont {Anke}}, \bibinfo
  {author} {\bibfnamefont {S.}~\bibnamefont {Berendts}}, \bibinfo {author}
  {\bibfnamefont {M.}~\bibnamefont {Lerch}}, \ and\ \bibinfo {author}
  {\bibfnamefont {T.}~\bibnamefont {Bredow}},\ }\href@noop {} {\bibfield
  {journal} {\bibinfo  {journal} {Solid State Sciences}\ }\textbf {\bibinfo
  {volume} {75}},\ \bibinfo {pages} {39} (\bibinfo {year} {2018})}\BibitemShut
  {NoStop}%
\bibitem [{\citenamefont {Hu}\ \emph {et~al.}(2019)\citenamefont {Hu},
  \citenamefont {He}, \citenamefont {Zhou}, \citenamefont {Li}, \citenamefont
  {Shen}, \citenamefont {Luo}, \citenamefont {Alsaedi}, \citenamefont {Hayat},
  \citenamefont {Zhou},\ and\ \citenamefont {Zou}}]{hu2019bivo}%
  \BibitemOpen
  \bibfield  {author} {\bibinfo {author} {\bibfnamefont {J.}~\bibnamefont
  {Hu}}, \bibinfo {author} {\bibfnamefont {H.}~\bibnamefont {He}}, \bibinfo
  {author} {\bibfnamefont {X.}~\bibnamefont {Zhou}}, \bibinfo {author}
  {\bibfnamefont {Z.}~\bibnamefont {Li}}, \bibinfo {author} {\bibfnamefont
  {Q.}~\bibnamefont {Shen}}, \bibinfo {author} {\bibfnamefont {W.}~\bibnamefont
  {Luo}}, \bibinfo {author} {\bibfnamefont {A.}~\bibnamefont {Alsaedi}},
  \bibinfo {author} {\bibfnamefont {T.}~\bibnamefont {Hayat}}, \bibinfo
  {author} {\bibfnamefont {Y.}~\bibnamefont {Zhou}}, \ and\ \bibinfo {author}
  {\bibfnamefont {Z.}~\bibnamefont {Zou}},\ }\href@noop {} {\bibfield
  {journal} {\bibinfo  {journal} {Chemical Communications}\ }\textbf {\bibinfo
  {volume} {55}},\ \bibinfo {pages} {5635} (\bibinfo {year}
  {2019})}\BibitemShut {NoStop}%
\bibitem [{\citenamefont {Yin}\ \emph {et~al.}(2011)\citenamefont {Yin},
  \citenamefont {Wei}, \citenamefont {Al-Jassim}, \citenamefont {Turner},\ and\
  \citenamefont {Yan}}]{yin2011doping}%
  \BibitemOpen
  \bibfield  {author} {\bibinfo {author} {\bibfnamefont {W.-J.}\ \bibnamefont
  {Yin}}, \bibinfo {author} {\bibfnamefont {S.-H.}\ \bibnamefont {Wei}},
  \bibinfo {author} {\bibfnamefont {M.~M.}\ \bibnamefont {Al-Jassim}}, \bibinfo
  {author} {\bibfnamefont {J.}~\bibnamefont {Turner}}, \ and\ \bibinfo {author}
  {\bibfnamefont {Y.}~\bibnamefont {Yan}},\ }\href@noop {} {\bibfield
  {journal} {\bibinfo  {journal} {Physical Review B}\ }\textbf {\bibinfo
  {volume} {83}},\ \bibinfo {pages} {155102} (\bibinfo {year}
  {2011})}\BibitemShut {NoStop}%
\bibitem [{\citenamefont {Zhao}\ \emph {et~al.}(2018)\citenamefont {Zhao},
  \citenamefont {Hu}, \citenamefont {Yao}, \citenamefont {Chen},\ and\
  \citenamefont {Chen}}]{zhao2018clarifying}%
  \BibitemOpen
  \bibfield  {author} {\bibinfo {author} {\bibfnamefont {X.}~\bibnamefont
  {Zhao}}, \bibinfo {author} {\bibfnamefont {J.}~\bibnamefont {Hu}}, \bibinfo
  {author} {\bibfnamefont {X.}~\bibnamefont {Yao}}, \bibinfo {author}
  {\bibfnamefont {S.}~\bibnamefont {Chen}}, \ and\ \bibinfo {author}
  {\bibfnamefont {Z.}~\bibnamefont {Chen}},\ }\href@noop {} {\bibfield
  {journal} {\bibinfo  {journal} {ACS Applied Energy Materials}\ }\textbf
  {\bibinfo {volume} {1}},\ \bibinfo {pages} {3410} (\bibinfo {year}
  {2018})}\BibitemShut {NoStop}%
\bibitem [{\citenamefont {Yang}\ \emph {et~al.}(2013)\citenamefont {Yang},
  \citenamefont {Kang}, \citenamefont {Sim}, \citenamefont {Lee}, \citenamefont
  {Lee}, \citenamefont {Koo}, \citenamefont {Nam},\ and\ \citenamefont
  {Joo}}]{yang2013new}%
  \BibitemOpen
  \bibfield  {author} {\bibinfo {author} {\bibfnamefont {T.-Y.}\ \bibnamefont
  {Yang}}, \bibinfo {author} {\bibfnamefont {H.-Y.}\ \bibnamefont {Kang}},
  \bibinfo {author} {\bibfnamefont {U.}~\bibnamefont {Sim}}, \bibinfo {author}
  {\bibfnamefont {Y.-J.}\ \bibnamefont {Lee}}, \bibinfo {author} {\bibfnamefont
  {J.-H.}\ \bibnamefont {Lee}}, \bibinfo {author} {\bibfnamefont
  {B.}~\bibnamefont {Koo}}, \bibinfo {author} {\bibfnamefont {K.~T.}\
  \bibnamefont {Nam}}, \ and\ \bibinfo {author} {\bibfnamefont {Y.-C.}\
  \bibnamefont {Joo}},\ }\href@noop {} {\bibfield  {journal} {\bibinfo
  {journal} {Physical Chemistry Chemical Physics}\ }\textbf {\bibinfo {volume}
  {15}},\ \bibinfo {pages} {2117} (\bibinfo {year} {2013})}\BibitemShut
  {NoStop}%
\bibitem [{\citenamefont {Liu}\ \emph {et~al.}(2020)\citenamefont {Liu},
  \citenamefont {Cui},\ and\ \citenamefont {Dupuis}}]{liu2020hole}%
  \BibitemOpen
  \bibfield  {author} {\bibinfo {author} {\bibfnamefont {T.}~\bibnamefont
  {Liu}}, \bibinfo {author} {\bibfnamefont {M.}~\bibnamefont {Cui}}, \ and\
  \bibinfo {author} {\bibfnamefont {M.}~\bibnamefont {Dupuis}},\ }\href@noop {}
  {\bibfield  {journal} {\bibinfo  {journal} {The Journal of Physical Chemistry
  C}\ }\textbf {\bibinfo {volume} {124}},\ \bibinfo {pages} {23038} (\bibinfo
  {year} {2020})}\BibitemShut {NoStop}%
\bibitem [{\citenamefont {Selim}\ \emph {et~al.}(2019)\citenamefont {Selim},
  \citenamefont {Pastor}, \citenamefont {Garc{\'\i}a-Tecedor}, \citenamefont
  {Morris}, \citenamefont {Franc{\`a}s}, \citenamefont {Sachs}, \citenamefont
  {Moss}, \citenamefont {Corby}, \citenamefont {Mesa}, \citenamefont {Gimenez},
  \citenamefont {Kafizas}, \citenamefont {Bakulin},\ and\ \citenamefont
  {Durrant}}]{selim2019impact}%
  \BibitemOpen
  \bibfield  {author} {\bibinfo {author} {\bibfnamefont {S.}~\bibnamefont
  {Selim}}, \bibinfo {author} {\bibfnamefont {E.}~\bibnamefont {Pastor}},
  \bibinfo {author} {\bibfnamefont {M.}~\bibnamefont {Garc{\'\i}a-Tecedor}},
  \bibinfo {author} {\bibfnamefont {M.~R.}\ \bibnamefont {Morris}}, \bibinfo
  {author} {\bibfnamefont {L.}~\bibnamefont {Franc{\`a}s}}, \bibinfo {author}
  {\bibfnamefont {M.}~\bibnamefont {Sachs}}, \bibinfo {author} {\bibfnamefont
  {B.}~\bibnamefont {Moss}}, \bibinfo {author} {\bibfnamefont {S.}~\bibnamefont
  {Corby}}, \bibinfo {author} {\bibfnamefont {C.~A.}\ \bibnamefont {Mesa}},
  \bibinfo {author} {\bibfnamefont {S.}~\bibnamefont {Gimenez}}, \bibinfo
  {author} {\bibfnamefont {A.}~\bibnamefont {Kafizas}}, \bibinfo {author}
  {\bibfnamefont {A.~A.}\ \bibnamefont {Bakulin}}, \ and\ \bibinfo {author}
  {\bibfnamefont {J.~R.}\ \bibnamefont {Durrant}},\ }\href@noop {} {\bibfield
  {journal} {\bibinfo  {journal} {Journal of the American Chemical Society}\
  }\textbf {\bibinfo {volume} {141}},\ \bibinfo {pages} {18791} (\bibinfo
  {year} {2019})}\BibitemShut {NoStop}%
\bibitem [{\citenamefont {Perdew}\ \emph {et~al.}(1996)\citenamefont {Perdew},
  \citenamefont {Burke},\ and\ \citenamefont {Ernzerhof}}]{Perdew1996}%
  \BibitemOpen
  \bibfield  {author} {\bibinfo {author} {\bibfnamefont {J.~P.}\ \bibnamefont
  {Perdew}}, \bibinfo {author} {\bibfnamefont {K.}~\bibnamefont {Burke}}, \
  and\ \bibinfo {author} {\bibfnamefont {M.}~\bibnamefont {Ernzerhof}},\ }\href
  {\doibase 10.1103/PhysRevLett.77.3865} {\bibfield  {journal} {\bibinfo
  {journal} {Physical Review Letters}\ }\textbf {\bibinfo {volume} {77}},\
  \bibinfo {pages} {3865} (\bibinfo {year} {1996})}\BibitemShut {NoStop}%
\bibitem [{\citenamefont {Giannozzi}\ \emph {et~al.}(2009)\citenamefont
  {Giannozzi} \emph {et~al.}}]{Giannozzi2009}%
  \BibitemOpen
  \bibfield  {author} {\bibinfo {author} {\bibnamefont {Giannozzi}} \emph
  {et~al.},\ }\href {\doibase 10.1088/0953-8984/21/39/395502} {\bibfield
  {journal} {\bibinfo  {journal} {Journal of Physics: Condensed Matter}\
  }\textbf {\bibinfo {volume} {21}},\ \bibinfo {pages} {395502} (\bibinfo
  {year} {2009})}\BibitemShut {NoStop}%
\bibitem [{\citenamefont {Deslippe}\ \emph {et~al.}(2012)\citenamefont
  {Deslippe}, \citenamefont {Samsonidze}, \citenamefont {Strubbe},
  \citenamefont {Jain}, \citenamefont {Cohen},\ and\ \citenamefont
  {Louie}}]{deslippe2012berkeleygw}%
  \BibitemOpen
  \bibfield  {author} {\bibinfo {author} {\bibfnamefont {J.}~\bibnamefont
  {Deslippe}}, \bibinfo {author} {\bibfnamefont {G.}~\bibnamefont
  {Samsonidze}}, \bibinfo {author} {\bibfnamefont {D.~A.}\ \bibnamefont
  {Strubbe}}, \bibinfo {author} {\bibfnamefont {M.}~\bibnamefont {Jain}},
  \bibinfo {author} {\bibfnamefont {M.~L.}\ \bibnamefont {Cohen}}, \ and\
  \bibinfo {author} {\bibfnamefont {S.~G.}\ \bibnamefont {Louie}},\ }\href@noop
  {} {\bibfield  {journal} {\bibinfo  {journal} {Computer Physics
  Communications}\ }\textbf {\bibinfo {volume} {183}},\ \bibinfo {pages} {1269}
  (\bibinfo {year} {2012})}\BibitemShut {NoStop}%
\bibitem [{\citenamefont {Hybertsen}\ and\ \citenamefont
  {Louie}(1986)}]{Hybertsen1986}%
  \BibitemOpen
  \bibfield  {author} {\bibinfo {author} {\bibfnamefont {M.~S.}\ \bibnamefont
  {Hybertsen}}\ and\ \bibinfo {author} {\bibfnamefont {S.~G.}\ \bibnamefont
  {Louie}},\ }\href {\doibase 10.1103/PhysRevB.34.5390} {\bibfield  {journal}
  {\bibinfo  {journal} {Physical Review B}\ }\textbf {\bibinfo {volume} {34}},\
  \bibinfo {pages} {5390} (\bibinfo {year} {1986})}\BibitemShut {NoStop}%
\bibitem [{\citenamefont {Rohlfing}\ and\ \citenamefont
  {Louie}(1998)}]{Rohlfing1998}%
  \BibitemOpen
  \bibfield  {author} {\bibinfo {author} {\bibfnamefont {M.}~\bibnamefont
  {Rohlfing}}\ and\ \bibinfo {author} {\bibfnamefont {S.~G.}\ \bibnamefont
  {Louie}},\ }\href {\doibase 10.1103/PhysRevLett.81.2312} {\bibfield
  {journal} {\bibinfo  {journal} {Physical Review Letters}\ }\textbf {\bibinfo
  {volume} {81}},\ \bibinfo {pages} {2312} (\bibinfo {year}
  {1998})}\BibitemShut {NoStop}%
\bibitem [{\citenamefont {Rohlfing}\ and\ \citenamefont
  {Louie}(2000)}]{rohlfing2000electron}%
  \BibitemOpen
  \bibfield  {author} {\bibinfo {author} {\bibfnamefont {M.}~\bibnamefont
  {Rohlfing}}\ and\ \bibinfo {author} {\bibfnamefont {S.~G.}\ \bibnamefont
  {Louie}},\ }\href@noop {} {\bibfield  {journal} {\bibinfo  {journal}
  {Physical Review B}\ }\textbf {\bibinfo {volume} {62}},\ \bibinfo {pages}
  {4927} (\bibinfo {year} {2000})}\BibitemShut {NoStop}%
\bibitem [{\citenamefont {Qiu}\ \emph {et~al.}(2016)\citenamefont {Qiu},
  \citenamefont {da~Jornada},\ and\ \citenamefont {Louie}}]{qiu2016screening}%
  \BibitemOpen
  \bibfield  {author} {\bibinfo {author} {\bibfnamefont {D.~Y.}\ \bibnamefont
  {Qiu}}, \bibinfo {author} {\bibfnamefont {F.~H.}\ \bibnamefont {da~Jornada}},
  \ and\ \bibinfo {author} {\bibfnamefont {S.~G.}\ \bibnamefont {Louie}},\
  }\href@noop {} {\bibfield  {journal} {\bibinfo  {journal} {Physical Review
  B}\ }\textbf {\bibinfo {volume} {93}},\ \bibinfo {pages} {235435} (\bibinfo
  {year} {2016})}\BibitemShut {NoStop}%
\bibitem [{\citenamefont {Stoughton}\ \emph {et~al.}(2013)\citenamefont
  {Stoughton}, \citenamefont {Showak}, \citenamefont {Mao}, \citenamefont
  {Koirala}, \citenamefont {Hillsberry}, \citenamefont {Sallis}, \citenamefont
  {Kourkoutis}, \citenamefont {Nguyen}, \citenamefont {Piper}, \citenamefont
  {Tenne}, \citenamefont {Podraza}, \citenamefont {Muller}, \citenamefont
  {Adamo},\ and\ \citenamefont {Schlom}}]{stoughton2013adsorption}%
  \BibitemOpen
  \bibfield  {author} {\bibinfo {author} {\bibfnamefont {S.}~\bibnamefont
  {Stoughton}}, \bibinfo {author} {\bibfnamefont {M.}~\bibnamefont {Showak}},
  \bibinfo {author} {\bibfnamefont {Q.}~\bibnamefont {Mao}}, \bibinfo {author}
  {\bibfnamefont {P.}~\bibnamefont {Koirala}}, \bibinfo {author} {\bibfnamefont
  {D.~A.}\ \bibnamefont {Hillsberry}}, \bibinfo {author} {\bibfnamefont
  {S.}~\bibnamefont {Sallis}}, \bibinfo {author} {\bibfnamefont {L.~F.}\
  \bibnamefont {Kourkoutis}}, \bibinfo {author} {\bibfnamefont
  {K.}~\bibnamefont {Nguyen}}, \bibinfo {author} {\bibfnamefont {L.~F.~J.}\
  \bibnamefont {Piper}}, \bibinfo {author} {\bibfnamefont {D.~A.}\ \bibnamefont
  {Tenne}}, \bibinfo {author} {\bibfnamefont {N.~J.}\ \bibnamefont {Podraza}},
  \bibinfo {author} {\bibfnamefont {D.~A.}\ \bibnamefont {Muller}}, \bibinfo
  {author} {\bibfnamefont {C.}~\bibnamefont {Adamo}}, \ and\ \bibinfo {author}
  {\bibfnamefont {D.~G.}\ \bibnamefont {Schlom}},\ }\href@noop {} {\bibfield
  {journal} {\bibinfo  {journal} {{APL} Materials}\ }\textbf {\bibinfo {volume}
  {1}},\ \bibinfo {pages} {042112} (\bibinfo {year} {2013})}\BibitemShut
  {NoStop}%
\bibitem [{\citenamefont {Cooper}\ \emph {et~al.}(2015)\citenamefont {Cooper},
  \citenamefont {Gul}, \citenamefont {Toma}, \citenamefont {Chen},
  \citenamefont {Liu}, \citenamefont {Guo}, \citenamefont {Ager}, \citenamefont
  {Yano},\ and\ \citenamefont {Sharp}}]{cooper2015indirect}%
  \BibitemOpen
  \bibfield  {author} {\bibinfo {author} {\bibfnamefont {J.~K.}\ \bibnamefont
  {Cooper}}, \bibinfo {author} {\bibfnamefont {S.}~\bibnamefont {Gul}},
  \bibinfo {author} {\bibfnamefont {F.~M.}\ \bibnamefont {Toma}}, \bibinfo
  {author} {\bibfnamefont {L.}~\bibnamefont {Chen}}, \bibinfo {author}
  {\bibfnamefont {Y.-S.}\ \bibnamefont {Liu}}, \bibinfo {author} {\bibfnamefont
  {J.}~\bibnamefont {Guo}}, \bibinfo {author} {\bibfnamefont {J.~W.}\
  \bibnamefont {Ager}}, \bibinfo {author} {\bibfnamefont {J.}~\bibnamefont
  {Yano}}, \ and\ \bibinfo {author} {\bibfnamefont {I.~D.}\ \bibnamefont
  {Sharp}},\ }\href@noop {} {\bibfield  {journal} {\bibinfo  {journal} {The
  Journal of Physical Chemistry C}\ }\textbf {\bibinfo {volume} {119}},\
  \bibinfo {pages} {2969} (\bibinfo {year} {2015})}\BibitemShut {NoStop}%
\bibitem [{\citenamefont {Payne}\ \emph {et~al.}(2011)\citenamefont {Payne},
  \citenamefont {Robinson}, \citenamefont {Egdell}, \citenamefont {Walsh},
  \citenamefont {McNulty}, \citenamefont {Smith},\ and\ \citenamefont
  {Piper}}]{payne2011nature}%
  \BibitemOpen
  \bibfield  {author} {\bibinfo {author} {\bibfnamefont {D.}~\bibnamefont
  {Payne}}, \bibinfo {author} {\bibfnamefont {M.}~\bibnamefont {Robinson}},
  \bibinfo {author} {\bibfnamefont {R.}~\bibnamefont {Egdell}}, \bibinfo
  {author} {\bibfnamefont {A.}~\bibnamefont {Walsh}}, \bibinfo {author}
  {\bibfnamefont {J.}~\bibnamefont {McNulty}}, \bibinfo {author} {\bibfnamefont
  {K.}~\bibnamefont {Smith}}, \ and\ \bibinfo {author} {\bibfnamefont
  {L.}~\bibnamefont {Piper}},\ }\href@noop {} {\bibfield  {journal} {\bibinfo
  {journal} {Applied Physics Letters}\ }\textbf {\bibinfo {volume} {98}},\
  \bibinfo {pages} {212110} (\bibinfo {year} {2011})}\BibitemShut {NoStop}%
\end{thebibliography}


\end{document}